\documentclass[12pt, nohyper]{JHEP3}
\usepackage{epsfig,amssymb,euscript}
\usepackage{amsmath,amscd}

\numberwithin{equation}{section}
\newcommand{\be}{\begin{eqnarray}}
\newcommand{\ee}{\end{eqnarray}}
\newcommand{\bea}{\begin{eqnarray}}
\newcommand{\eea}{\end{eqnarray}}
\newcommand{\ba}{\begin{array}}
\newcommand{\ea}{\end{array}}
\newcommand{\nn}{\nonumber \\}

\newcommand{\bR}{\mathbb{R}}
\newcommand{\bC}{\mathbb{C}}

\def\bo{{\bar{1}}}

\catcode`\@=11
\newcount\hour
\newcount\minute
\newtoks\amorpm \hour=\time\divide\hour by 60\minute
=\time{\multiply\hour by 60 \global\advance\minute by-\hour}
\edef\standardtime{{\ifnum\hour<12 \global\amorpm={am}%
        \else\global\amorpm={pm}\advance\hour by-12 \fi
        \ifnum\hour=0 \hour=12 \fi
        \number\hour:\ifnum\minute<10
        0\fi\number\minute\the\amorpm}}
\edef\militarytime{\number\hour:\ifnum\minute<10
0\fi\number\minute}
\def\draftlabel#1{{\@bsphack\if@filesw {\let\thepage\relax
   \xdef\@gtempa{\write\@auxout{\string
      \newlabel{#1}{{\@currentlabel}{\thepage}}}}}\@gtempa
   \if@nobreak \ifvmode\nobreak\fi\fi\fi\@esphack}
        \gdef\@eqnlabel{#1}}
\def\@eqnlabel{}
\def\@vacuum{}
\def\marginnote#1{}
\def\draftmarginnote#1{\marginpar{\raggedright\scriptsize\tt#1}}
\def\draft{
        \pagestyle{plain}
        \overfullrule=2pt
        \oddsidemargin -.1truein
        \def\@oddhead{\sl \phantom{\today\quad\militarytime} \hfil
        \smash{\Large\sl DRAFT} \hfil \today\quad\militarytime}
        \let\@evenhead\@oddhead
        \let\label=\draftlabel
        \let\marginnote=\draftmarginnote
        \def\ps@empty{\let\@mkboth\@gobbletwo
        \def\@oddfoot{\hfil \smash{\Large\sl DRAFT} \hfil}
        \let\@evenfoot\@oddhead}
        \def\@eqnnum{(\theequation)\rlap{\kern\marginparsep\tt\@eqnlabel}%
        \global\let\@eqnlabel\@vacuum}  }

\renewcommand{\theequation}{\thesection.\arabic{equation}}

\def\appendix#1{\addtocounter{section}{1}\setcounter{equation}{0}
\renewcommand{\thesection}{\Alph{section}}
\section*{Appendix \thesection\protect\indent \parbox[t]{11.15cm}{#1}}
\addcontentsline{toc}{section}{Appendix \thesection\ \ \ #1}}

\title{Null Half-Supersymmetric Solutions in Five-Dimensional Supergravity}

\author{Jai Grover and Jan B. Gutowski  \\ DAMTP, Centre for Mathematical Sciences\\
University of Cambridge\\
Wilberforce Road, Cambridge, CB3 0WA, UK\\ E-mail: \email{ J.Grover@damtp.cam.ac.uk},  \email{J.B.Gutowski@damtp.cam.ac.uk}}

\author{Wafic Sabra \\ Centre for Advanced Mathematical Sciences and Physics Department
\\ American University of Beirut \\ Lebanon \\ E-mail : \email{ws00@aub.edu.lb}}

\abstract{We classify half-supersymmetric solutions of
gauged $N=2$, $D=5$ supergravity coupled to an arbitrary number of abelian
vector multiplets for which all of the Killing spinors generate null Killing vectors.
We show that there are four classes of solutions, and in each class we find the metric,
scalars and gauge field strengths. When the scalar manifold is symmetric,
the solutions correspond to a class of
local near horizon geometries recently found by Kunduri and Lucietti. }

\begin{document}



\section{Introduction}

The classification of supergravity solutions has many applications in
string theory. Such classifications have been used recently to
construct new  black hole and black
ring solutions. Furthermore, the classifications can also be used to
prove non-existence theorems in several supergravity theories, whereby
solutions preserving certain proportions of supersymmetry are excluded.
Recently, a partial classification of solutions of $N=2$, $D=5$
was constructed \cite{sabragut2007}. Solutions with four
linearly independent Killing spinors for which at least
two generate a timelike Killing vector were completely classified.
In this paper we complete the classification of half-supersymmetric
solutions of $N=2$, $D=5$ supergravity by considering the case
when all four Killing spinors generate null Killing vectors.

There are a number of interesting supersymmetric solutions in
$N=2$, $D=5$ supergravity. Supersymmetric solutions can in
principle preserve $1/4$, $1/2$, $3/4$ or the maximal proportion of
supersymmetry. Examples of $1/4$ supersymmetric solutions are for
instance the regular asymptotically $AdS_5$ black holes found in
\cite{gutreall2004} and later generalized in
\cite{popelu2005} and \cite{reallluc2006}. $1/4$-supersymmetric
string solutions have also been constructed in \cite{chamssabra99}
and \cite{sabraklemm2000c}. In \cite{gutgaunt2003} a classification of
all $1/4$-supersymmetric solutions of minimal gauged $ N = 2, D = 5$
supergravity was performed, this was later extended to a classification
of $1/4$-supersymmetric solutions of a more general $N=2$, $D=5$
gauged supergravity, coupled to an arbitrary number of abelian
vector multiplets. Examples of $1/2$-supersymmetric solutions are the domain wall
solutions in \cite{sabrklemm2003}, as well as the solutions given in
\cite{sabrchamsed1998}, \cite{sabrklemm2000a} and
\cite{sabrklemm2000b} which correspond to black holes without
regular horizons. The regular asymptotically $AdS_5$ black
holes also undergo supersymmetry enhancement in their near-horizon limit
from $1/4$ to $1/2$ supersymmetry, as do the black string solutions in
\cite{sabraklemm2000c}. In \cite{grovergut2006}, it was shown that
all $3/4$-supersymmetric solutions must be locally $AdS_5$, although
globally there exist discrete qoutients of $AdS_5$ which are $3/4$-supersymmetric
\cite{figgut2007}. The unique maximally supersymmetric solution is $AdS_5$.

In order to investigate half-supersymmetric null solutions we will make
use of the spinorial geometry method. This method was first used
to classify solutions of supergravity theories in ten and eleven
dimensions \cite{papadgran2005a}, \cite{papadgran2005b}.
The first step of such analysis is to write the spinors
of the theory as differential forms.
The gauge symmetries of the supergravity theories are then used to
simplify the spinors as much as possible.
By choosing an appropriate basis, the Killing spinor equations
(or their integrability conditions) are written as a linear system.
This linear system can be solved to
express the fluxes of the theory in terms of the geometry and to find the conditions on the
geometry imposed by supersymmetry.
These methods have also been
 particularly useful in classifying solutions which preserve
very large amounts of supersymmetry; for example in
\cite{papadgran2006b} it has been shown that all solutions preserving
$29/32$, $30/32$ or $31/32$ of the supersymmetry are in fact maximally
supersymmetric. We also remark that the spinorial geometry
method has been used to classify solutions of $N=2$, $D=4$ supergravity;
see for example \cite{roestklemm2007}.

The plan of this paper is as follows. In Section 2, we review some of
the properties of five dimensional gauged supergravity coupled to
abelian vector multiplets. In Section 3, we show how spinors of the
theory can be written as differential forms, and introduce an
adapted basis in the forms suitable for defining null Killing
spinors. We then use the $Spin(4,1)$ gauge freedom present in the
theory to reduce one null Killing spinor into a particularly simple
canonical form, and the residual symmetry present to place the other
null spinor into one of two forms.  In Section 4, we summarize the
constraints imposed by solutions preserving $1/4$ of the
supersymmetry.
In Sections 5 and 6 we derive constraints on the spacetime geometry,
the gauge field strengths and the scalars obtained from the Killing spinor
equations. A number of different cases are examined in detail, corresponding
to the various different ways in which the Killing spinors can be simplified
using gauge transformations. In section 7, we present a self-contained
summary of the metrics, scalars and gauge field strengths for all of
these half-supersymmetric solutions, together with an interpretation
of these solutions. Finally, in
Appendix A, we show that the integrability conditions of the Killing
spinor equations together with the Bianchi identity are sufficient to
ensure that the Einstein, gauge and scalar equations hold automatically.
In Appendix B, we present a detailed derivation of the linear system obtained
from the Killing spinor equations for half-supersymmetric null solutions.

\section{$N=2, D=5$ Supergravity}

We begin by briefly reviewing some aspects of $N=2$, $D=5$ gauged
supergravity coupled to $n$ abelian vector multiplets. The bosonic
action of this theory is~\cite{gunaydin:85}

\begin{eqnarray}
\label{bosact} S = {\frac{1 }{16 \pi G}} \int \big( (-{}^5 R + 2
\chi^2 {\mathcal{V}}) \star 1
-Q_{IJ} F^I \wedge *F^J +Q_{IJ} dX^I \wedge \star dX^J  \notag \\
-{\frac{1 }{6}} C_{IJK} F^I \wedge F^J \wedge A^K \big)
\end{eqnarray}

where $I,J,K$ take values $1, \ldots ,n$ and $F^I=dA^I$. $C_{IJK}$
are constants that are symmetric on $IJK$; we will assume that
$Q_{IJ}$ is invertible, with inverse $Q^{IJ}$. The metric has
signature $(+,-,-,-,-)$.

The $X^I$ are scalars which are constrained via
\begin{equation}  \label{eqn:conda}
{\frac{1 }{6}}C_{IJK} X^I X^J X^K=1 \,.
\end{equation}

We may regard the $X^I$ as being functions of $n-1$ unconstrained
scalars $\phi^a$. It is convenient to define

\begin{equation}
X_I \equiv {\frac{1 }{6}}C_{IJK} X^J X^K
\end{equation}
so that the condition~({\ref{eqn:conda}}) becomes
\begin{equation}
X_I X^I =1\,.
\end{equation}
In addition, the coupling $Q_{IJ}$ depends on the scalars via
\begin{equation}
Q_{IJ} = {\frac{9 }{2}} X_I X_J -{\frac{1 }{2}}C_{IJK} X^K
\end{equation}
so in particular
\begin{equation}
Q_{IJ} X^J = {\frac{3 }{2}} X_I\,, \qquad Q_{IJ} \partial_a X^J =
-{\frac{3 }{2}} \partial_a X_I\,.
\end{equation}

The scalar potential can be written as,

\begin{equation}
{\mathcal{V}} = 9 V_I V_J (X^I X^J - {\frac{1 }{2}} Q^{IJ}) \ ,
\end{equation}
where $V_I$ are constants.

For a bosonic background to be supersymmetric there must be a spinor
$\epsilon$ for
which the supersymmetry variations of the gravitino and the
superpartners of the scalars vanish. We shall investigate the
properties of these spinors in greater detail in the next section.
The gravitino Killing spinor equation is

\begin{equation}
\label{eqn:grav} \big( \partial_{\mu} +
{1\over4}\omega_{\mu}{}^{\rho\sigma}\Gamma_{\rho\sigma}
-{3i\chi\over2} A_{\mu} +{i\chi\over2}V_IX^I\Gamma_{\mu}
-{3\over4}H_{\mu}{}^{\rho}\Gamma_{\rho} +
{1\over8}\Gamma_{\mu}H^{\rho\sigma}\Gamma_{\rho\sigma} \big)\epsilon
= 0 \ ,
\end{equation}

where $\epsilon$ is a Dirac spinor. The algebraic Killing spinor
equations associated with the variation of the scalar superpartners
are

\begin{equation}
\label{eqn:newdil} \big(4i\chi(X^IV_JX^J - {3\over2}Q^{IJ}V_J) +
2\partial^\mu X^I \Gamma_{\mu} - (F^{I\mu\nu} -
X^IH^{\mu\nu})\Gamma_{\mu\nu}\big)\epsilon = 0 \ .
\end{equation}

where we define $ H = X_I F^I , A = V_IA^I$. We shall refer to
({\ref{eqn:newdil}}) as the dilatino Killing spinor equation. We
also require that the bosonic background should satisfy the
Einstein, gauge field and scalar field equations obtained from the
action ({\ref{bosact}}) and analyse these in Appendix A.

\section{Spinors in Five Dimensions}

Following \cite{lawson, wang, harvey}, the space of Dirac spinors in
five dimensions is the space of complexified forms on $\bR^2$,
$\Delta = \Lambda^* (\bR^2) \otimes \bC$. A generic spinor $\eta$ can
therefore be written as
\begin{equation}
\eta = \lambda 1 + \mu^i e^i + \sigma e^{12} \ ,
\end{equation}
where $e^1$, $e^2$ are 1-forms on $\bR^2$, and $i=1,2$ for complex functions
$\lambda, \mu^i$ and $\sigma$.
The action of $\gamma$-matrices on these forms is given by
\begin{eqnarray}
\gamma_i &=& i (e^i \wedge + i_{e^i})\ ,
\\
\gamma_{i+2} &=& -e^i \wedge +i_{e^i} \ , \end{eqnarray}
for $i=1,2$. $\gamma_0$ is defined by
\begin{equation}
\gamma_0 = \gamma_{1234} \ ,
\end{equation}
and satisfies
\begin{equation}
\gamma_0 1 = 1 , \quad \gamma_0 e^{12} = e^{12} , \quad \gamma_0 e^i = -e^i \ \ i=1,2 \ .
\end{equation}
The charge conjugation operator $C$ is defined by
\begin{equation}
C 1 = -e^{12}, \quad C e^{12} = 1 \quad C e^i = - \epsilon_{ij} e^j \ \ i=1,2
\end{equation}
where $\epsilon_{ij}=\epsilon^{ij}$ is antisymmetric with $\epsilon_{12}=1$.
We also note the useful identity
\begin{equation}
(\gamma_M)^* = - \gamma_0 C \gamma_M \gamma_0 C \  .
\end{equation}

\subsection{The Null Basis}

To work in a basis adapted to describing solutions with Killing spinors
which generate null Killing vectors, define

\begin{eqnarray} \Gamma_{\pm}&=& {1 \over \sqrt{2}}(\gamma_0 \mp \gamma_3) \ , \nn
\Gamma_1 &=& {1 \over \sqrt{2}}(\gamma_2 -i \gamma_4) = \sqrt{2} i
e^2 \wedge  \ , \nn \Gamma_\bo &=& {1 \over \sqrt{2}}(\gamma_2+i\gamma_4)
= \sqrt{2} i i_{e^2}\ , \nn \Gamma_2 &=& \gamma_1 \ .\end{eqnarray}

We then define a basis for the Dirac spinors $\Delta$ by

\begin{equation} \psi^1_\pm = 1 \pm e^1, \qquad \qquad \psi^\bo_\pm = e^{12}
\mp e^2 \ .\end{equation}

Note that $\psi^1_\pm$ is {\it not} the complex conjugate of $\psi^\bo_\pm$.

Then it is straightforward to show that

\begin{eqnarray} \Gamma_\pm \psi^\alpha_\pm &=& 0 \nn \Gamma_\pm
\psi^\alpha_\mp &=& \sqrt{2} \psi^\alpha_\pm \nn \Gamma_\alpha
\psi^\beta_\pm &=& \mp \sqrt{2} i \delta_\alpha^\beta
\psi^{\bar{\beta}}_\pm \nn \Gamma_2 \psi^1_\pm &=& \pm i \psi^1_\pm
\nn \Gamma_2 \psi^\bo_\pm &=& \mp i \psi^\bo_\pm  \ , \end{eqnarray}

where $\alpha, \beta =1, \bo$ and we use the index convention that
$\psi^{\bar{\bo}}_\pm = \psi^1_\pm$.

A generic spinor can then be written as

\begin{equation}
\eta = \lambda^\alpha_+ \psi^\alpha_+ + \lambda^\alpha_-
\psi^\alpha_- \ ,
\end{equation}

where there is summation over $\alpha=1, \bo$. Note that the $\lambda^\alpha_\pm$
are in general complex and $\lambda^1_\pm$ is {\it not} the complex conjugate
of $\lambda^\bo_\pm$.

The metric has vielbein ${\bf{e}}^+, {\bf{e}}^-, {\bf{e}}^1, {\bf{e}}^\bo, {\bf{e}}^2$,
where ${\bf{e}}^{\pm},
{\bf{e}}^2$ are real, and ${\bf{e}}^1, {\bf{e}}^\bo$ are complex conjugate, and

\begin{equation}
ds^2 = 2 {\bf{e}}^+ {\bf{e}}^- -2 {\bf{e}}^1 {\bf{e}}^\bo - ({\bf{e}}^2)^2 \ .
\end{equation}

Now note that on writing the Dirac spinor $\eta$ as
$\eta=\eta^1+i\eta^2$, where $\eta^a$ are symplectic Majorana
spinors, we find

\begin{equation} B(\eta^1, \eta^2) = {1 \over 2} B(\gamma_0 C \eta^*, \eta) =
-{1 \over 2} \langle \gamma_0 \eta, \eta \rangle \ .\end{equation}

Hence the nullity condition $B(\eta^1, \eta^2)=0$ can be rewritten
in the null basis as

\begin{equation} \label{nullity} \lambda^1_+ (\lambda^1_-)^* + (\lambda^1_+)^*
\lambda^1_- +\lambda^\bo_+ (\lambda^\bo_-)^* + (\lambda^\bo_+)^*
\lambda^\bo_-  = 0 \ . \end{equation}

To proceed further, note that
\begin{equation}
e^{x \gamma_{03} + y \gamma_{24}} (1+e^1) = e^{x-iy} (1+e^1) \ ,
\end{equation}
for $x,y \in \bR$, and it is also convenient to define
\begin{equation}
T_1 = \gamma_{01}+\gamma_{13}, \quad T_2 = \gamma_{02}+\gamma_{23},
\quad T_3 = \gamma_{04} - \gamma_{34} \ ,
\end{equation}
which satisfy
\begin{equation}
T_i \psi^\alpha_+ =0 \ ,
\end{equation}
for $\alpha=1, \bo$, and also
\begin{equation}
T_1 \psi^1_- = -2i \psi^1_+, \quad T_1 \psi^\bo_- = 2i \psi^\bo_+ \
,
\end{equation}
\begin{equation}
T_2 \psi^1_- = 2i \psi^\bo_+, \quad T_2 \psi^\bo_- = 2i \psi^1_+ \ ,
\end{equation}
\begin{equation}
 T_3 \psi^1_- = -2 \psi^\bo_+, \quad T_3 \psi^\bo_- = 2
\psi^1_+ \ .\end{equation}

Note that gauge transformations of the form $e^{xT_1+yT_2+zT_3}$
for $x,y,z \in \bR$ map

\begin{eqnarray} \label{gaugetransf} \lambda^\alpha_- &\rightarrow&
\lambda^\alpha_- \nn \lambda^1_+ &\rightarrow& \lambda^1_+ -ix
\lambda^1_- + (z+iy)\lambda^\bo_- \nn \lambda^\bo_+ &\rightarrow&
\lambda^\bo_+ +ix \lambda^\bo_- + (iy-z)\lambda^1_- \
.\end{eqnarray}

Clearly these leave $1 + e^1$ invariant. We therefore adopt the
following approach. Using the $Spin(4,1)$ gauge freedom, we can
choose without loss of generality the first Killing spinor to be

\begin{equation} \epsilon = \psi^1_+  \ . \end{equation}

The gauge transformations $e^{xT_1+yT_2+zT_3}$ leave $\epsilon$
invariant. The second Killing spinor of the form
\be \eta = \lambda^\alpha_+ \psi^\alpha_+ + \lambda^\alpha_-
\psi^\alpha_- \ee
where $\lambda^\alpha_\pm$ satisfy ({\ref{nullity}}) can then be
simplified by using the gauge transformations $e^{xT_1+yT_2+zT_3}$.

In particular, we note that we can make use of the gauge transformations to
set either $\lambda_+^{\alpha} = 0$, or $\lambda_-^{\alpha} = 0 $.
To see this, let us first assume that $\lambda_-^1 \neq 0$ and
$\lambda_-^{\bo} \neq 0 $. Then we can use ({\ref{gaugetransf}}) to
set $\lambda_+^1 = 0$ by imposing

\begin{equation} (z+iy)\lambda_-^{\bar{1}} - ix\lambda_-^1 = \lambda_+^1 \ .\end{equation}

This fixes $z,y$ in the $\lambda_+^{\bar{1}}$ transformation

\begin{eqnarray} \lambda_+^\bo &\rightarrow& \lambda_+^\bo + ix\lambda_-^\bo -
{\lambda_-^1\over\lambda_-^\bo{}^*}\big(\lambda_+^1{}^* -
ix\lambda_-^1{}^*\big) \nn &=&
{1\over\lambda_-^\bo{}^*}\big(\lambda_+^\bo\lambda_-^\bo{}^* -
\lambda_-^1\lambda_+^1{}^* +ix(\lambda_-^\bo\lambda_-^\bo{}^* +
\lambda_-^1\lambda_-^1{}^*)\big) \ . \end{eqnarray}

We can fix $x$ here such that the term in brackets is real; then we
find

\begin{eqnarray} \lambda_+^1 &=& 0 \nn \lambda_+^\bo &=&
\mu\lambda_-^\bo \ ,
\end{eqnarray}

with $\mu \in \bR$. To proceed further we use this result together
with the nullity condition ({\ref{nullity}}) to find

\begin{equation} 2\mu\lambda_-^\bo\lambda_-^\bo{}^* = 0 \ . \end{equation}

This implies that $\mu = 0 $. Alternatively, we have the case where
$\lambda_-^1 = 0, \lambda_-^\bo \neq 0$. Here we can use $y,z$ in
({\ref{gaugetransf}}) to set $\lambda_+^1 = 0$. This sets

\begin{eqnarray} \lambda_+^\bo &\rightarrow& \lambda_+^\bo + ix\lambda_-^\bo \nn
&=& \lambda_-^\bo(ix+{\lambda_+^\bo\over\lambda_-^\bo}) \ .
\end{eqnarray}

Here $x$ can be chosen to set the term in brackets to be real, so
that once again we have

\begin{eqnarray} \lambda_+^1 &=& 0 \nn \lambda_+^\bo &=& \mu\lambda_-^\bo =
0 \ , \end{eqnarray}

where we set $\mu = 0$ using the nullity condition as before. The
case $\lambda_-^1 \neq 0, \lambda_-^\bo = 0 $ proceeds analogously.

\section{Quarter-Supersymmetric Null Solutions}

In appendix B we arrive at the general linear system following from
the dilatino and gravitino equations acting on a spinor $\epsilon =
\lambda_+^1 \psi_+^1 + \lambda_+^{\bo} \psi_+^{\bo} + \lambda_-^1
\psi_-^1 + \lambda_-^{\bo} \psi_-^{\bo}$. Restricting to the case
$\epsilon = \psi^1_+$ we find

\begin{equation} F^{I}_{+-}  = 0 \ , \end{equation}

\begin{equation} \label{F11bar} F^{I}_{1\bar{1}} = -i(-\partial_2X^I +
2\chi(X^IV_JX^J - {3\over2}Q^{IJ}V_J)) + X^IH_{1\bar{1}} \ ,
\end{equation}

\begin{equation} \partial_+ X^I = 0 \ , \end{equation}

\begin{equation} F^{I}_{+2} = 0 \ , \end{equation}

\begin{equation} F^{I}_{+\bo}  = 0 \ , \end{equation}

\begin{equation} \label{F12} F^{I}_{\bo2} = i\partial_{\bo}X^I + X^IH_{\bo2} \ .\end{equation}

{}Further constraints on the spin connection obtained from the gravitino equation acting on
$\epsilon = \psi_+^1$ are

\begin{equation} \omega_{+,+-} = \omega_{+,+2} = \omega_{+,+\bo} = \omega_{-,+-}
= \omega_{1,+2} = \omega_{\bo,+\bo} = 0 ,\end{equation}

and

\begin{equation} \omega_{\bo,\bo2} =\omega_{2,+2} =  \omega_{+,\bo2} =
\omega_{2,+1} = \omega_{1,+\bo} =0 \ , \end{equation}

as well as

\begin{equation} \omega_{\bo,+-} + \omega_{-,+\bo} = 0 \ , \end{equation}

\begin{equation} -\omega_{\bo,+-} + {1\over2}\omega_{2,\bo2} = 0 \ , \end{equation}

\begin{equation} -2i\omega_{-,+2} + i\omega_{1,\bo2} + 3i\chi V_IX^I = 0 \ , \end{equation}

\begin{equation} \omega_{2,+-} + \omega_{-,+2} = 0 \ , \end{equation}

We also find

\begin{equation} H_{+1} = H_{+2} = H_{+-} = 0 \ , \end{equation}

\begin{equation} H_{-\bo} = -{2i\over3}\omega_{-,\bo2} \ , \end{equation}

\begin{equation} H_{-2} = 2\chi A_- - {2i\over3}\omega_{-,1\bar{1}} \ , \end{equation}

\begin{equation} H_{\bo2} = 2i\omega_{\bo,+-} \ , \end{equation}

\begin{equation} H_{1\bar{1}} = -{2i\over3}\omega_{-,+2}
-{2i\over3}\omega_{1,\bo2} \ , \end{equation}

where the gauge potential has the following components constrained

\begin{equation} \chi A_{\bar{1}} = {2i\over3}\omega_{\bo,+-} +
{i\over3}\omega_{\bo,1\bo} \ , \end{equation}

\begin{equation} \chi A_2 = {i\over3}\omega_{2,1\bo} \ , \end{equation}

\begin{equation}  \chi A_+ = {i\over3}\omega_{+,1\bo} \ . \end{equation}

To proceed to half-supersymmetric solutions, we incorporate these
constraints into the full linear system in Appendix B and consider
two cases in which either $\lambda_+^{\alpha} = 0$ or
$\lambda_-^{\alpha} = 0$.

\section{Solutions with $\lambda_+^{\alpha} = 0$}

For this class of solutions, we set $\lambda_+^{\alpha} = 0$
for $\alpha = 1, \bo$, in the components of the dilatino and
gravitino Killing spinor equations, with the resulting linear system
presented in Appendix B. For a non-trivial solution to ({\ref{21}}),
and ({\ref{22}}) to exist, we require

\begin{eqnarray} 2(-\chi A_- -{i\over3}\omega_{-1\bar{1}} -
{i\over2}\omega_{2,-2})(-\chi A_- -{i\over3}\omega_{-,1\bo} -
{i\over2}\omega_{2,-2}) \nn + (-\omega_{2,-\bo} +
{1\over3}\omega_{-,\bo2})(-\omega_{2,-1} + {1\over3}\omega_{-,12}) =
0 \ , \end{eqnarray}

which implies that

\begin{eqnarray} \chi A_-  = {i\over3}\omega_{-,1\bo} \ , \end{eqnarray}

\begin{eqnarray} \omega_{2,-2} = 0 \ , \end{eqnarray}

\begin{eqnarray} \omega_{2,-\bo} = {1\over3}\omega_{-,\bo2} \ . \end{eqnarray}

Using ({\ref{11}}), and ({\ref{1bar2}}) we require

\begin{eqnarray} {1\over2}(\omega_{-,12} + \omega_{1,-2})(\omega_{-,\bo2} +
\omega_{\bo,-2}) + (\omega_{1,-1})(\omega_{\bo,-\bo}) = 0 \ ,
\end{eqnarray}

which implies that

\begin{eqnarray} \omega_{-,12} = -\omega_{1,-2} \ , \end{eqnarray}

\begin{eqnarray} \omega_{1,-1} = 0 \ . \end{eqnarray}

We can also use ({\ref{-1}}), and ({\ref{-2}}) finding that

\begin{eqnarray} (\omega_{-,-2})(\omega_{-,-2}) +
(\omega_{-,-\bo})(\omega_{-,-1}) = 0 \ , \end{eqnarray}

so that

\begin{eqnarray} \omega_{-,-2} = 0 \ , \end{eqnarray}

\begin{eqnarray} \omega_{-,-1} = 0 \ . \end{eqnarray}

{}From ({\ref{12}}), and ({\ref{1bar1}})

\begin{eqnarray} (\omega_{1,-\bo})(\omega_{\bo,-1}) +
{8\over9}(\omega_{1,-2})(\omega_{\bo,-2}) = 0 \ , \end{eqnarray}

from which we see that

\begin{eqnarray} \omega_{1,-\bo} = 0 \ , \end{eqnarray}

\begin{eqnarray} \omega_{1,-2} =  \omega_{-,12} = \omega_{2,-1} =
0 \ . \end{eqnarray}

Using the dilatino equations ({\ref{d1}}) and ({\ref{d2}}), we
require that

\begin{eqnarray} 8\big(\partial_-X^I - i(F^{I}_{-2} - X^IH_{-2})\big)
\big(\partial_-X^J + i(F^{J}_{-2} - X^JH_{-2})\big) \nn +
16(F^{I}_{-\bo} - X^IH_{-\bo})(F^{J}_{-1} - X^JH_{-1}) = 0 \ ,
\end{eqnarray}

so that, upon contracting with $Q_{IJ}$

\begin{eqnarray} F^{I}_{-\bo} = X^IH_{-\bo} = 0  \ , \end{eqnarray}

\begin{eqnarray} F^I_{-2} =  X^IH_{-2} = 0 \ , \end{eqnarray}

\begin{eqnarray} \partial_- X^I = 0 \ . \end{eqnarray}

Within the case $\lambda_+^{\alpha} = 0$ there are three sub-cases
to consider. Here either $(\lambda^1_- \neq 0, \lambda^{\bar{1}}_-
\neq 0)$, or $(\lambda^1_- = 0, \lambda^{\bo}_- \neq 0)$, or
$(\lambda^1_- \neq 0, \lambda^{\bo}_- = 0)$.

\subsection{Solutions with $\lambda^1_- \neq 0$ and
$\lambda^{\bar{1}}_- \neq 0$}

Suppose first that $\lambda^1_- \neq 0$ and $\lambda^{\bar{1}}_-
\neq 0$. Then note that the $U(1) \times Spin(4,1)$ gauge
transformation of the type $e^{ig\mu} e^{g \mu \gamma_{24}}$ for
$\mu \in \bR, g \in \bR$ which acts on spinors via

\begin{eqnarray} \label{gauge2} \psi^1_\pm \rightarrow e^{ig\mu} e^{g \mu \gamma_{24}} \psi^1_\pm &=&
\psi^1_\pm \nn  \psi^{\bar{1}}_\pm  \rightarrow e^{ig \mu} e^{g \mu \gamma_{24}} \psi^{\bar{1}}_\pm
&=& e^{2ig \mu} \psi^{\bar{1}}_\pm \ , \end{eqnarray}

leaves $\epsilon=\psi^1_+$ invariant, and transforms $\eta$ as

\begin{eqnarray} \eta &\rightarrow& \lambda^1_-\psi^1_- +
e^{2ig\mu}\lambda^{\bo}_-\psi^{\bo}_-  =
(\lambda^1_-)'\psi^1_- + (\lambda^{\bo}_-)'\psi^{\bo}_- \ .
\end{eqnarray}

Define

\begin{equation} g =
i\log{\lambda^1_-\lambda^{\bar{1}}_-\over(\lambda^1_-\lambda^{\bar{1}}_-)^*}
\ .
\end{equation}

Then we find that

\begin{equation}
{(\lambda^1_-)'(\lambda^{\bar{1}}_-)'\over((\lambda^1_-)'(\lambda^{\bar{1}}_-)')^*}
=
\big({\lambda^1_-\lambda^{\bar{1}}_-\over(\lambda^1_-\lambda^{\bar{1}}_-)^*}\big)^{1-4\mu}
\ .
\end{equation}

Hence, for $\mu ={1\over4}$, and dropping the primes, we have

\begin{equation} \label{lamrel1}
{\lambda^1_-\lambda^{\bar{1}}_-\over(\lambda^1_-\lambda^{\bar{1}}_-)^*}
=1 \ . \end{equation}

Now, observe that

\begin{equation}
\partial_+ g = -2i \omega_{+,1 \bar{1}}, \qquad \partial_- g = -2i \omega_{-,1
\bar{1}} \ , \end{equation}

so that, working in this gauge, we can take without loss of
generality

\begin{equation} \omega_{+,1 \bar{1}}=\omega_{-,1 \bar{1}} =0  \ . \end{equation}

Note in particular that in this gauge

\begin{equation}
\partial_+ \lambda^{\bar{1}}_- = \partial_- \lambda^{\bar{1}}_- = \partial_+ \lambda^1_- =\partial_- \lambda^1_-
=0 \ . \end{equation}

To proceed, we investigate several integrability conditions. In
particular, requiring that $\nabla_{[+} \nabla_{-]}
\lambda^{\bar{1}}_-=0$ imposes the constraint

\begin{eqnarray}
(\omega_{+,- \bar{1}} - \omega_{-,+\bar{1}}) (-\omega_{1,1
\bar{1}}\lambda^{\bar{1}}_- - \sqrt{2} \omega_{1, \bar{1}2}
\lambda^1_-) -(\omega_{+,-1}-\omega_{-,+1}) \omega_{\bar{1},1
\bar{1}} \lambda^{\bar{1}}_- \nn
+(\omega_{+,-2}-\omega_{-,+2})(-\sqrt{2} \omega_{2,\bar{1}2}
\lambda^1_- +(-\omega_{2,1 \bar{1}}-{1 \over 3}
\omega_{\bar{1},12}+{2 \over 3}\omega_{-,+2}) \lambda^{\bar{1}}_- )
=0 \ ,
\end{eqnarray}

and requiring that $\nabla_{[+} \nabla_{-]} \lambda^1_-=0$ imposes the constraint

\begin{eqnarray}
\label{auxint1} \big( {2 \sqrt{2} \over
3}(\omega_{+,-1}-\omega_{-,+1})(\omega_{1, \bar{1}2}+\omega_{-,+2})
+\sqrt{2}(\omega_{+,-2}-\omega_{-,+2}) \omega_{2,12} \big)
\lambda^{\bar{1}}_- \nn +2
(\omega_{+,-1}\omega_{-,+\bar{1}}-\omega_{+,-\bar{1}}\omega_{-,+\bar{1}})
\lambda^1_- =0 \ .
\end{eqnarray}

Next, the conditions $\nabla_{[\pm} \nabla_{B]} \lambda^1_- = \nabla_{[\pm} \nabla_{B]} \lambda^{\bar{1}}_-=0$
for $B=1, \bar{1}, 2$ impose the constraints

\begin{eqnarray}
\partial_\pm \omega_{2,12} = \partial_\pm \omega_{2,1 \bar{1}} = \partial_\pm \omega_{1,+-} =
\partial_\pm \omega_{1, 1 \bar{1}} = \partial_\pm \omega_{1, \bar{1}2} = \partial_\pm \omega_{-,+2} =0
\ .
\end{eqnarray}

Now note that in the gauge for which $A_+ = A_- = 0$, we have
\begin{equation}
\chi A = {i \over 3} \omega_{2, 1 \bar{1}} {\bf{e}}^2 + {i \over
3}(2 \omega_{\bar{1},+-}+\omega_{\bar{1}, 1 \bar{1}})
{\bf{e}}^{\bar{1}} -{i \over 3}(2 \omega_{1,+-}-\omega_{1,1
\bar{1}}) {\bf{e}}^1 \ .
\end{equation}

The integrability condition $d(\chi A)_{+-}=0$ then implies that

\begin{eqnarray} \label{dA+-} (2 \omega_{\bar{1},+-}+\omega_{\bar{1},1
\bar{1}})(-\omega_{+,-1}+\omega_{-,+1}) +(-2
\omega_{1,+-}+\omega_{1,1 \bar{1}})
(-\omega_{+,-\bar{1}}+\omega_{-,+\bar{1}}) \nn + \omega_{2,1
\bar{1}} (-\omega_{+,-2}+\omega_{-,+2})
 =0 \ .
\end{eqnarray}

Note also that ({\ref{+1}}) and ({\ref{+2}}) can be rewritten as

\begin{equation} \label{new+1} {1 \over \sqrt{2}}(\omega_{+,-2}+\omega_{-,+2})
\lambda^1_- -(\omega_{+,-1}+\omega_{-,+1}) \lambda^{\bar{1}}_- =0 \
,
\end{equation}

and

\begin{equation} \label{new+2} -(\omega_{+,-\bar{1}}+\omega_{-,+\bar{1}})
\lambda^1_- +(-{1 \over \sqrt{2}} \omega_{+,-2} +{1 \over 3
\sqrt{2}}\omega_{-,+2}-{\sqrt{2} \over 3}\omega_{1, \bar{1} 2})
\lambda^{\bar{1}}_- = 0 \ . \end{equation}

Next note that the component of the Bianchi identity $X_I dF^I_{+-2}=0$ implies that

\begin{equation}
\label{intsimp1} \omega_{-,+1} \omega_{+,- \bar{1}}-\omega_{-,+
\bar{1}} \omega_{+,-1} = 0 \ , \end{equation}

and substituting this into ({\ref{auxint1}}) we find

\begin{equation} \label{intsimp2}
{1\over3}(\omega_{+,-1}-\omega_{-,+1})(\omega_{1,\bar{1}
2}-\omega_{-,+2}) -\omega_{-,+1}(\omega_{+,-2} -\omega_{-,+2}) = 0 \
.
\end{equation}

Using these identities we obtain the constraints

\begin{equation}
{1 \over 2}((\omega_{+,-2})^2-(\omega_{-,+2})^2) +
\omega_{-,+1}\omega_{-,+\bar{1}} -\omega_{+,-\bar{1}} \omega_{+,-1}
= 0 \ , \end{equation}

\begin{equation} \label{intsimp3} (\omega_{+,-2} +
\omega_{-,+2})\big((\omega_{+,-\bo} - \omega_{-,+\bo})\lambda^1_- +
{1\over\sqrt{2}}(\omega_{+,-2} - \omega_{-,+2})\lambda^{\bo}_-\big)
= 0 \ . \end{equation}

We now find cases according as to whether $(\omega_{+,-2} +
\omega_{-,+2})$ vanishes. First suppose $(\omega_{+,-2} +
\omega_{-,+2}) = 0 $. Then ({\ref{new+1}}) and ({\ref{new+2}}) imply
that

\begin{equation} 2\omega_{-,+2} - \omega_{1,\bo2} = 3\chi V_IX^I = 0 \ . \end{equation}

Contracting ({\ref{d4}}) with $V_I$ then implies that $Q^{IJ}V_IV_J
= 0$. As $Q^{IJ}$ is positive definite this is a contradiction.

We are then led to take $(\omega_{+,-2} + \omega_{-,+2}) \neq 0 $.
In this case we have the constraint

\begin{equation} \label{constraint} (\omega_{+,-\bo} -
\omega_{-,+\bo})\lambda^1_- + {1\over\sqrt{2}}(\omega_{+,-2} -
\omega_{-,+2})\lambda^{\bo}_- = 0 \ . \end{equation}

Further simplifications can be made by going back to our gauge
transformations ({\ref{gauge2}}). Requiring $\partial_1 g = 0$
implies that

\begin{equation} \sqrt{2} \omega_{1,1 \bar{1}} \lambda^{\bar{1}}_- = -\omega_{1,
\bar{1}2} \lambda^1_- \ , \end{equation}

when taken together with ({\ref{intsimp3}}). Similarly, $\partial_2
g = 0 $ can be shown to require that

\begin{equation} \chi A_2 = \omega_{2,1\bo} = 0 \ . \end{equation}

These conditions are sufficient to show that

\begin{equation} d ({\lambda^{\bo}_-\over(\lambda^{\bo}_-)^*}) = 0 \ , \end{equation}

and hence from ({\ref{lamrel1}}) that

\begin{equation} d ({\lambda^{1}_-\over(\lambda^{1}_-)^*}) = 0 \ . \end{equation}

Then, by making use of the $U(1) \times Spin(4,1)$ gauge transformation
of the type $e^{i \theta_1} e^{\theta_2 \gamma_{24}}$ for constant
$\theta_1, \theta_2 \in \bR$, we can set, without loss of generality
\begin{equation} {\lambda^{\bo}_-\over(\lambda^{\bo}_-)^*} =
{\lambda^{1}_-\over(\lambda^{1}_-)^*} =1 \ . \end{equation}
This gauge transformation multiplies $\psi^1_+$ by a phase, however
as this phase is constant, it does not alter the constraints obtained
in the analysis of the quarter-supersymmetric solutions.

Using these results, we find the following constraints remain on the
spatial derivatives of the $\lambda$'s;

\begin{equation} \partial_1\lambda^1_- = -2\omega_{-,+1}\lambda^1_-  \ , \end{equation}

\begin{equation} \partial_1\lambda^{\bo}_- = {-1\over\sqrt{2}}\omega_{1,\bo2}
\lambda^1_-  \ , \end{equation}

\begin{equation} \partial_2\lambda^1_- =
-2\sqrt{2}\omega_{-,+1}\lambda^{\bo}_- \ ,
\end{equation}

\begin{equation} \partial_2\lambda^{\bo}_- = -\omega_{1,\bo2}\lambda^{\bo}_- \ . \end{equation}

To proceed we note that

\begin{equation} \label{v}  V = ({(\lambda^1_-)^2 + (\lambda^{\bo}_-)^2 \over
\sqrt{2}}){\bf{e}}^+  \ , \end{equation}

\begin{equation} \label{w} W = {\bf{e}}^-  \ , \end{equation}

are Killing vectors of the theory. We can find an additional Killing
vector $U$, as

\begin{equation} \label{u} U = [V,W] = c_1Y  \ , \end{equation}

where $Y$ is defined by

\begin{equation} \label{y} Y = \lambda^{\bo}_-({\bf{e}}^1 + {\bf{e}}^{\bo})
-\sqrt{2}\lambda^1_- {\bf{e}}^2 \ , \end{equation}

and $c_1$ by

\begin{equation} \label{c1} c_1 = \omega_{-,+2}\lambda^1_-
-\sqrt{2}\omega_{-,+1}\lambda^{\bo}_- \ . \end{equation}

As $Y$ can also be shown to be Killing we find that $c_1$ must be a
constant.

We define a vector orthogonal to $V$, $W$, and $Y$ as

\begin{equation} \label{z} Z = \lambda^{1}_-({\bf{e}}^1+{\bf{e}}^{\bo}) +
\sqrt{2}\lambda^{\bo}_- {\bf{e}}^2  \ , \end{equation}

and a vector orthogonal to $V$, $W$, $Y$ and $Z$, as

\begin{equation} \label{x} X = i\lambda^{\bo}_-({\bf{e}}^1 - {\bf{e}}^{\bo}) \ , \end{equation}

where $X$ can also be shown to be Killing. Furthermore we find

\begin{equation} \label{dV} dV =  {1\over f}(c_1Y +c_2Z) \wedge V  \ , \end{equation}

\begin{equation} \label{dW} dW = {1\over f}(-c_1Y + c_2Z) \wedge W  \ ,  \end{equation}

\begin{equation} \label{dX} dX = -{\omega_{1,\bo2}\sqrt{2}\over
\lambda^{\bo}_-} Z\wedge X  \ , \end{equation}

\begin{equation} \label{dY} dY = -{2\sqrt{2}c_1\over f}V \wedge W +{c_2\over f}Z
\wedge Y  \ , \end{equation}

\begin{equation} \label{dZ} dZ = 0  \ , \end{equation}

\begin{equation} \label{dlambdaone} d\lambda^1_- = -2\omega_{-,+1}Z \ , \end{equation}

\begin{equation} \label{dlambdabar} d\lambda^{\bo}_- =
{-1\over\sqrt{2}}\omega_{1,\bo2}Z  \ . \end{equation}

Here $c_2$ and $f$ are given by

\begin{equation} \label{c2} c_2 = \sqrt{2}\omega_{-,+1}\lambda^1_- +
\omega_{-,+2}\lambda^\bo_- \ , \end{equation}

\begin{equation} \label{f} f = ({(\lambda^1_-)^2 + (\lambda^{\bo}_-)^2 \over
\sqrt{2}}) \ , \end{equation}

\begin{equation} \label{df} df = c_2Z \ , \end{equation}

and $c_1, c_2$, and $f$ are related by

\begin{equation} \label{c1c21} -c_1\lambda^\bo_- + c_2\lambda^1_- =
2\omega_{-,+1}f \ , \end{equation}

\begin{equation} \label{c1c22} c_1\lambda^1_- + c_2\lambda^{\bo}_- =
\sqrt{2}\omega_{-,+2}f \ , \end{equation}

\begin{equation} \label{c1spin}(\omega_{-,+1} - \omega_{+,-1})f =
-c_1\lambda^{\bo}_-  \ , \end{equation}

\begin{equation} \label{c2spin} (\omega_{-,+2} - \omega_{+,-2})f =
\sqrt{2}c_1\lambda^1_-  \ .\end{equation}

{}From ({\ref{new+1}}), ({\ref{new+2}}), and ({\ref{constraint}}) we
find that

\begin{equation} c_2 = \chi V_I X^I \lambda^{\bo}_-  \ , \end{equation}

which together with ({\ref{df}}) implies that

\begin{equation} \label{partialzf} \partial_z f = \chi V_I X^I\lambda^{\bo}_-  \ .\end{equation}

In addition, ({\ref{dlambdaone}}) and ({\ref{c1c21}}) can be
combined in the following way

\begin{equation} \label{flambdaone} d(f\lambda^1_-) = c_1\lambda^{\bo}_- Z
 \ . \end{equation}

The forms $V$, $W$, $X$, $Y$, and $Z$, can be expressed in terms of
coordinates as

\begin{equation} V = f_1 dv  \ , \end{equation}

\begin{equation} W = f_2 dw  \ , \end{equation}

\begin{equation} X = f_3 dx  \ , \end{equation}

\begin{equation} Y = f(dy + \beta)  \ , \end{equation}

\begin{equation} Z = dz \ . \end{equation}

The coordinate derivatives of the scalars ({\ref{d3}}) and
({\ref{d4}}) are

\begin{equation} \label{dyXI} \partial_y X_I = 0 \ , \end{equation}

\begin{equation} \label{dzXI} -\partial_z X_I = {\chi\over f}(X_IV_JX^J -
V_I)\lambda^{\bo}_-  \ , \end{equation}

which implies that

\begin{equation} \label{dXI} dX_I = -{\chi\over f}(X_IV_JX^J -
V_I)\lambda^{\bo}_- Z \ . \end{equation}

The functions $f_1,f_2,f_3$ and the form $\beta$ can be constrained,
upon comparison with ({\ref{dV} - \ref{dZ}}), by

\begin{equation} d\log{f_1} = c_1(dy + \beta) + d\log{f} + G dv \ , \end{equation}

\begin{equation} d\log{f_2} = -c_1(dy + \beta) + d\log{f} + H dw  \ , \end{equation}

\begin{equation} d\log{f_3} = d\log{(\lambda^{\bo}_-)^2}  \ , \end{equation}

\begin{equation} \label{dbeta} d\beta = {-2\sqrt{2}c_1f_1f_2\over f^2} dv \wedge
dw \ . \end{equation}

We can rewrite these as

\begin{equation} \label{f1f2} d\log{{f_1f_2\over f^2}} = G dv + H dw  \ , \end{equation}

\begin{equation} \label{f1/f2} d\log{{f_1\over f_2}} = 2c_1(dy + \beta) + G dv - H
dw  \ , \end{equation}

\begin{equation} f_3 = c_3(\lambda^{\bo}_-)^2  \ . \end{equation}

for $c_3$ a non-zero constant. Taking the exterior derivative of
({\ref{f1/f2}}) and ({\ref{f1f2}}) we find respectively

\begin{equation} \label{dbeta2} 2 c_1 d\beta = (\partial_v H + \partial_w G)dv
\wedge dw  \ , \end{equation}

\begin{equation} (-\partial_w G + \partial_v H)dv \wedge dw = 0  \ . \end{equation}

Upon comparing ({\ref{dbeta2}}) with ({\ref{dbeta}}) we see that $G$
and $H$ have only a $v$ and $w$ dependance

\begin{equation} \partial_v H = \partial_w G = {-2\sqrt{2}(c_1)^2 f_1 f_2\over
f^2}  \ , \end{equation}

and satisfy

\begin{equation} \label{h} \partial_w\partial_v H = H\partial_v H  \ , \end{equation}

\begin{equation} \label{g} \partial_v\partial_w G = G\partial_w G  \ . \end{equation}

The field strength $F^I$ takes the form

\begin{equation} F^I = F^I_{12} {\bf{e}}^1 \wedge {\bf{e}}^2 + F^I_{\bo2} {\bf{e}}^{\bo}
 \wedge {\bf{e}}^2 +
F^I_{1\bo} {\bf{e}}^1 \wedge {\bf{e}}^{\bo}  , \end{equation}

with non-zero components, $ F^I_{12}, F^I_{\bo2}, F^I_{1\bo}$, given
by ({\ref{F11bar}}) and ({\ref{F12}}). These can be expressed in
terms of the scalars using the scalar derivatives ({\ref{dX}}),
together with ({\ref{dlambdaone}}) and ({\ref{dA+-}}), as

\begin{equation} F^I = d({ X^I\lambda^1_- c_3 dx\over\sqrt{2}}) \ . \end{equation}

The scalar derivatives ({\ref{dXI}}) can in turn be put into the
form

\begin{equation} \label{intXI} d(fX_I) = \chi V_I\lambda^{\bo}_- Z \ . \end{equation}

using ({\ref{partialzf}}). To proceed we need to consider two cases
depending on whether $c_1$ vanishes or not.

\subsubsection{Solutions with $ c_1 = 0 $}

In the case that $c_1 = 0 $, ({\ref{flambdaone}}) reduces to

\begin{equation} d(f\lambda^1_-) = 0  \ , \end{equation}

so that

\begin{equation} f\lambda^1_- = c_4  \ , \end{equation}

for non-zero constant $c_4$. Here $f$ is implicitly related to the
scalars via the relation

\begin{equation} \partial_z (fX_I) = \chi V_I (\sqrt{2}f - {c_4^2\over
f^2})^{1\over2} \ . \end{equation}

We further find in this case that

\begin{equation} d \log{f_1 \over f} = G dv  \ , \end{equation}

\begin{equation} d\log{f_2\over f} = H dw  \ , \end{equation}

and that the metric is given by

\begin{equation} ds^2 = 2f(z)({f_1f_2\over f^2})dv dw -
{(c_3\lambda^{\bo}_-)^2\over 2}dx^2 - {1\over 2\sqrt{2}f}dz^2 -
{f\over 2\sqrt{2}}dy^2 \ . \end{equation}

where
\begin{equation}
\label{nnonconstxx}
 \lambda^{\bo}_- = \big(\sqrt{2}f - ({c_4 \over f})^2 \big)^{1\over
2} \ .
\end{equation}

Moreover, as $G$ and ${f_1\over f}$ can be seen to be functions
of $v$, and $H$ and ${f_2\over f}$ are functions of $w$, we find
that, for $c_1 = 0$, the 2-manifold given by, $ ^{(2)}ds^2 =
2({f_1f_2\over f^2})dv dw  $, is flat.

\subsubsection{Solutions with $ c_1 \neq 0 $}

On the other hand, if $c_1 \neq 0 $ then ({\ref{intXI}}), together
with ({\ref{flambdaone}}), can be explicitly integrated up to

\begin{equation} X_I = {1\over c_1}({K_I \over f} + \chi V_I \lambda^1_- )  \ , \end{equation}

with $K_I$ constant. The metric in our coordinates is now given,
more generally, by

\begin{equation} ds^2 = 2f(z)({f_1f_2\over f^2})dv dw -
{(c_3\lambda^{\bo}_-)^2\over 2}dx^2 - {1\over2\sqrt{2}f}dz^2 -
{f\over2\sqrt{2}}(dy + \beta)^2 \ . \end{equation}

In this case we can relate the function ${f_1f_2\over f^2}$ to the
Ricci scalar for the 2-manifold with metric, $ ^{(2)}ds^2 =
2{f_1f_2\over f^2}dv dw $. The Ricci scalar is given by

\begin{eqnarray} ^{(2)}R &=& {-2\over({-\partial_v
H\over2\sqrt{2}c_1{}^2})^3}({-1\over2\sqrt{2}c_1{}^2})^2\big(\partial_v
H\partial_v\partial_w\partial_v H - \partial_v\partial_v
H\partial_w\partial_v H \big) \nn &=& 4\sqrt{2}(c^1)^2 \
,\end{eqnarray}

where we have made use of ({\ref{h}}). This manifold is then found
to be $AdS_2$. We can also make a gauge transformation $\beta
\rightarrow \beta + d\log{{\tilde{f_1}\over \tilde{f_2}}}$, to
eliminate the $x$ and $z$ dependance of $\beta$. The $y$ dependance
of $f_1, f_2$ can further be expressed as

\begin{equation} f_1 = \tilde{f_1}\exp{(c_1y)}  \ , \end{equation}

\begin{equation} f_2 = \tilde{f_2}\exp{(-c_1y)}  \ , \end{equation}

so that ({\ref{f1/f2}}) reduces to

\begin{equation} \beta = H dw - G dv  \ , \end{equation}

where $\beta$ is only a function of $v$ and $w$.

\subsection{Solutions with $\lambda^1_- = 0$ and
$\lambda^{\bar{1}}_- \neq 0$}

The $\lambda^\bo_-$ derivatives are

\begin{equation} \partial_+\lambda^\bo_- = -\omega_{+,1\bo}\lambda^\bo_-  \ , \end{equation}

\begin{equation} \partial_-\lambda^\bo_- = (-\omega_{-,1\bo} +
\omega_{\bo,-1})\lambda^{\bo}_-  , \end{equation}

\begin{equation} \partial_1 \lambda^\bo_- = -\omega_{1,1\bo}\lambda^\bo_-  \ , \end{equation}

\begin{equation} \partial_\bo \lambda^\bo_- = -\omega_{\bo,1\bo}\lambda^\bo_-  \ , \end{equation}

\begin{equation} \partial_2 \lambda^\bo_- = (\omega_{-,+2} -
\omega_{2,1\bo})\lambda^\bo_-  \ , \end{equation}

and the other non-zero components of the spin connection are related
by

\begin{equation} \omega_{-,+2} = \omega_{+,-2} = -\omega_{1,\bo2} = \chi V_I
X^I \ ,
\end{equation}

\begin{equation} \omega_{\bo,-1} = -{1\over2}\omega_{2,-2} \ . \end{equation}

We find for the scalars

\begin{equation} \label{dscalar2} dX^I = 2\chi(X^IV_JX^J - {3\over2}Q^{IJ}V_J)
e^2  \ , \end{equation}

and gauge potential

\begin{equation} \chi A_- = {i\over3}\omega_{-,1\bo} - i\omega_{\bo,-1} \ .
\end{equation}

We can use a gauge transformation as in ({\ref{gauge2}}), taking

\begin{equation} \psi^{\bar{1}}_\pm \rightarrow e^{2ig' \mu} \psi^{\bar{1}}_\pm  \ , \end{equation}

to set $\lambda^{\bo}_- \in \bR $. As a result we find

\begin{equation} \omega_{+,1\bo} = \omega_{-,1\bo} = \omega_{1,1\bo} =
\omega_{2,1\bo} = 0  \ , \end{equation}

and

\begin{equation} d\lambda^{\bo}_- = \omega_{-,+2}\lambda^{\bo}_- {\bf{e}}^2  \ . \end{equation}

The field strength $F^I$ vanishes in this case. We find closed forms

\begin{equation} V =  {\bf{e}}^+  \ , \end{equation}

\begin{equation} W = h^{-1}{\bf{e}}^-  \ , \end{equation}

\begin{equation} X = (\sqrt{2}h)^{-{1\over 2}} ({\bf{e}}^1 + {\bf{e}}^{\bo})  \ , \end{equation}

\begin{equation} Y = i (\sqrt{2}h)^{-{1\over 2}} ({\bf{e}}^1 - {\bf{e}}^{\bo})  \ ,  \end{equation}

\begin{equation} Z = {\bf{e}}^2  , \end{equation}

where $h = (\lambda^{\bo}_-)^2$. Then specify a coordinate basis

\begin{equation} V = dv, W = dw, X = dx, Y = dy, Z = dz \ . \end{equation}

In this basis

\begin{equation} dh = 2\chi V_I X^I h dz  \ , \end{equation}

so that upon comparison with ({\ref{dscalar2}}), we find that

\begin{equation} \partial_z (h X_I) = 2\chi V_I h  \ . \end{equation}

The metric is given by

\begin{equation} ds^2 = h ( 2 dv dw -  dx^2 -  dy^2) - dz^2 \ . \end{equation}

\subsection{Solutions with $\lambda^1_- \neq 0$ and
$\lambda^{\bar{1}}_- = 0$}

The $\lambda^1_-$ derivatives vanish in this case

\begin{equation} d\lambda^1_- = 0 \ . \end{equation}

The following components of the spin connection vanish

\begin{equation}  \omega_{-,+1} = \omega_{+,-1} = \omega_{1,\bo2} = 0 \ , \end{equation}

\begin{equation}  \omega_{1,-1} = \omega_{2,-1} = \omega_{-,-2} = \omega_{-,-1} =
\omega_{-,12} = \omega_{1,-2} = 0 \ , \end{equation}

and we have

\begin{equation} \omega_{-,+2} = -\omega_{+,-2} \ , \end{equation}

\begin{equation} \omega_{\bo,-1} = -{1\over2}\omega_{2,-2} = 0 \ . \end{equation}

We also find that the scalars are constant

\begin{equation} dX^I = 0  \ , \end{equation}

and for the gauge potential

\begin{equation} \chi A =  {i\over3}\omega_{+,1\bo}{\bf{e}}^+ +
{i\over3}\omega_{-,1\bo}{\bf{e}}^- + {i\over3}\omega_{1,1\bo}{\bf{e}}^1 +
{i\over3}\omega_{\bo,1\bo}{\bf{e}}^{\bo} + {i\over3}\omega_{2,1\bo}{\bf{e}}^2 \ .
\end{equation}

We can integrate up the scalars, in the process defining a constant
$c$ by

\begin{equation} \chi V_I X^I = c = {2\over3}\omega_{-,+2} \ ,  \end{equation}

with $X^I = q^I$. The field strengths have non-vanishing component

\begin{equation} F^I_{1\bo} = 3i\chi (-X^I V_J X^J + Q^{IJ}V_J)  \ , \end{equation}

which are therefore also constants. We can contract this with $\chi
V_I$, to find

\begin{equation} \label{FdA} F = \chi V_I F^I = i k {\bf{e}}^1 \wedge {\bf{e}}^{\bo} \ , \end{equation}

with constant $ k = -3\big( c^2 - \chi^2 Q^{IJ}V_I V_J) \big) $.

Taking the exterior derivative of the basis forms, one obtains

\begin{equation} \label{2d+} d{\bf{e}}^+  = -3c {\bf{e}}^2 \wedge {\bf{e}}^+ \ , \end{equation}

\begin{equation} \label{2d-} d{\bf{e}}^- = 3c {\bf{e}}^2 \wedge {\bf{e}}^-  \ , \end{equation}

\begin{equation} \label{2d1} d{\bf{e}}^1 = 3i\chi A \wedge {\bf{e}}^1  \ , \end{equation}

\begin{equation} \label{2d1bar} d{\bf{e}}^{\bo} = -3i\chi A \wedge {\bf{e}}^{\bo}  \ , \end{equation}

\begin{equation} \label{2d2} d{\bf{e}}^2 = 3c {\bf{e}}^+ \wedge {\bf{e}}^-  \ . \end{equation}

Coordinates can be introduced for ${\bf{e}}^+$ and $ {\bf{e}}^- $ as

\begin{equation} \label{vv2x} {\bf{e}}^+ = g_1 dv \ , \end{equation}

\begin{equation} \label{ww2x} {\bf{e}}^- = g_2 dw \ . \end{equation}

Comparing ({\ref{2d+}}) and ({\ref{2d-}}) with ({\ref{vv2x}}),
({\ref{ww2x}}),  we find

\begin{equation} \label{g1} d\log{g_1} = -3c{\bf{e}}^2 + 3c\alpha_1 dv \ , \end{equation}

\begin{equation} \label{g2} d\log{g_2} = 3c{\bf{e}}^2 - 3c\alpha_2 dw \ , \end{equation}

for some real functions $\alpha_1, \alpha_2$. These can be rewritten
as

\begin{equation} \label{g1g2} d\log{g_1g_2} = 3c(\alpha_1 dv - \alpha_2 dw) \ , \end{equation}

\begin{equation} \label{g1overg2} d\log{g_1\over g_2} = -6c {\bf{e}}^2 + 3c(\alpha_1 dv
+ \alpha_2 dw) \ . \end{equation}

Then ({\ref{g1overg2}}) defines ${\bf{e}}^2$ implicitly to be

\begin{equation} \label{zalpha} {\bf{e}}^2 = dz + {1\over 2}(\alpha_1 dv + \alpha_2 dw)
 \ , \end{equation}

where we define the coordinate $z$, such that, $dz =
{-1\over6c}d\log{g_1\over g_2}$. Next we can introduce complex
coordinates for ${\bf{e}}^1, {\bf{e}}^{\bo}$ as

\begin{equation} \label{l} {\bf{e}}^1 = s d\ell  \ , \end{equation}

\begin{equation} \label{lbar} {\bf{e}}^{\bo} = \bar{s} d\bar{\ell}  \ , \end{equation}

where $s = re^{i\theta}$, and $d\ell = dx + idy$. Then

\begin{equation} \label{ds} d\log{s} + q d\ell = 3i\chi A  \ , \end{equation}

\begin{equation} \label{dsbar} d\log{\bar{s}} + \bar{q} d\bar{\ell} = -3i \chi
A \ ,
\end{equation}

upon comparison with ({\ref{2d1}}) and ({\ref{2d1bar}}). Here $q$ is
a complex function $q = q_1 + i q_2$. These expressions can in turn
be rewritten as

\begin{equation} \label{ssbar} d\log{s\bar{s}} = - q d\ell - \bar{q}d\bar{\ell}
 \ , \end{equation}

\begin{equation} \label{soversbar} d\log{s\over\bar{s}} = 6i\chi A +
\bar{q} d\bar{\ell} - q d\ell  \ . \end{equation}

({\ref{soversbar}}) implicitly defines $A$, up to a gauge
transformation, as

\begin{equation} \chi A = {1\over 3} (q_2 dx + q_1 dy) \ . \end{equation}

With these coordinates, the metric takes the form

\begin{equation} ds^2 = 2g_1 g_2 dv dw - (dz + {\alpha \over 2})^2 - 2r^2(dx^2 + dy^2)  \ , \end{equation}

where $\alpha = \alpha_1 dv + \alpha_2 dw $. We can proceed to investigate the
curvature of  the 3-manifold with metric

\be \label{adsiii} ^{(3)}ds^2 = 2g_1g_2 dv dw - (dz + {\alpha \over
2})^2 \ . \ee

To do this we take the exterior derivative of ({\ref{g1g2}}) and
({\ref{g1overg2}})

\begin{equation} \label{a1a2} d \alpha_1 \wedge dv - d \alpha_2 \wedge dw = 0  \ , \end{equation}

\begin{equation} \label{a1} d{\bf{e}}^2 = {1\over2} ( d\alpha_1 \wedge dv + d\alpha_2
\wedge dw ) \ . \end{equation}

These constraints, together with ({\ref{2d2}}) imply

\begin{equation} \partial_v \alpha_2 =  -\partial_w \alpha_1 = 3c g_1 g_2 \ , \end{equation}

and that $\alpha_1 = \alpha_1 (v,w), \alpha_2 = \alpha_2 (v,w)$.
Substituting this back into the expression ({\ref{g1g2}}) for
$g_1g_2$ , we see that

\begin{equation} \label{da2} {d(\partial_v \alpha_2)\over
\partial_v \alpha_2} = 3c (\alpha_1 dv - \alpha_2 dw) \ . \end{equation}

Next we note that the 2-manifold with metric, $ ^{(2)}ds^2 = 2g_1g_2
dv dw $ is $AdS_2$ with Ricci scalar $18c^2$, and that $\alpha$ is
related to the volume form for this manifold by $d\alpha = 6c \ {\rm
dvol }(AdS_2)$. It then follows that the 3-manifold with metric
({\ref{adsiii}}) is $AdS_3$ (written as a fibration over $AdS_2$),
with Ricci scalar \be {}^{(3)}R = {27c^2\over2} \ . \ee

We can, in a similar manner,  compute the
Ricci scalar for the 2-manifold with metric $ ^{(2)}ds^2 = 2s\bar{s}
d\ell d\bar{\ell}  = 2 r^2 d\ell d\bar{\ell} $.

Taking the exterior derivative of ({\ref{ssbar}}) and
({\ref{soversbar}}) provides

\begin{equation} dq \wedge d\ell + d\bar{q} \wedge d\bar{\ell} = 0 \ , \end{equation}

\begin{equation} 6i\chi dA = dq \wedge d\ell - d\bar{q} \wedge d\bar{\ell} \  . \end{equation}

Given that $F = \chi dA $ we can compare this with ({\ref{FdA}}), to
find

\begin{equation} \partial_{\bar{\ell}}q = 3k r^2 \ , \end{equation}

where $ r^2 = s\bar{s}$. If we substitute this back into the
expression ({\ref{ssbar}}) for $s\bar{s}$ , we find that

\begin{equation} \label{q} {d(\partial_{\bar{\ell}}q)\over
\partial_{\bar{\ell}}q} = -q d\ell - \bar{q} d\bar{\ell} \ . \end{equation}

The Ricci scalar is given by (making use of ({\ref{q}}))

\begin{eqnarray} ^{(2)}R &=& {-2\over({1\over
3k}\partial_{\bar{\ell}}q)^3}({1\over
3k})^2\big(\partial_{\bar{\ell}}q\partial_{\ell}\partial_{\bar{\ell}}\partial_{\bar{\ell}}
q -
\partial_{\bar{\ell}}\partial_{\bar{\ell}}q\partial_{\ell}\partial_{\bar{\ell}} q \big) \nn &=&
6k\ .  \end{eqnarray}

The 2-manifold is then $\mathbb{H}^2$, $\bR^2$, or $S^2$
according as to whether the constant $k = -3(c^2 - \chi^2 Q^{IJ}V_I
V_J)$ is negative, vanishing, or positive respectively.

\section{Solutions with $ \lambda_-^{\alpha} = 0 $}

In the case where $\lambda_-^ {\alpha} = 0$ for $\alpha = 1, \bo$,
we find for the dilatino equations

\begin{equation} 8i\chi(X^IV_JX^J - {3\over2}Q^{IJ}V_J)\lambda_+^{\bar{1}} =0  \ . \end{equation}

For the gravitino equations, in the $+$ direction

\begin{equation} \partial_+\lambda_+^1 = 0  \ , \end{equation}

\begin{equation} \partial_{+}\lambda_+^{\bar{1}}
+\omega_{+,1\bar{1}}\lambda_+^{\bar{1}}= 0  \ .  \end{equation}

In the $-$ direction

\begin{equation} \partial_-\lambda_+^1 = 0   \ , \end{equation}

\begin{equation} (\partial_- -3i\chi A_-)\lambda_+^{\bar{1}}
+\omega_{-,1\bar{1}}\lambda_+^{\bar{1}} = 0  \ , \end{equation}

\begin{equation}  \sqrt{2}i\chi V_IX^I\lambda_+^{\bar{1}} = 0  \ .  \end{equation}

In the $1$ direction

\begin{equation} \partial_1\lambda_+^1  = 0  \ , \end{equation}

\begin{equation} \partial_1\lambda_+^{\bar{1}}  + \omega_{1,1\bar{1}}
\lambda_+^{\bar{1}} - 2\omega_{1,+-}\lambda_+^{\bar{1}}  = 0
 \ .  \end{equation}

In the $\bar{1}$ direction

\begin{equation} \partial_{\bar{1}}\lambda_+^1  +
\chi\sqrt{2}V_IX^I\lambda_+^{\bar{1}} = 0  \ , \end{equation}

\begin{equation} \partial_{\bar{1}}\lambda_+^{\bar{1}}
+2\omega_{\bo,+-}\lambda_+^{\bar{1}}  +
\omega_{\bo,1\bar{1}}\lambda_+^{\bar{1}} = 0 \ . \end{equation}

In the $2$ direction

\begin{equation} \partial_2\lambda_+^1  = 0  \ , \end{equation}

\begin{equation} \partial_2\lambda_+^{\bar{1}} + \chi V_IX^I\lambda_+^{\bar{1}} +
\omega_{2,1\bar{1}}\lambda_+^{\bar{1}}  = 0 \ . \end{equation}

These constraints imply that $\lambda^{\bar{1}}_+ = 0$ and that
$\lambda_+^1$ is constant. Hence these solutions are only $1/4$
supersymmetric.

\section{Summary of Results}

In this paper we examined half supersymmetric solutions of gauged
$N = 2, D = 5$ supergravity coupled to an arbitrary number of abelian
vector multiplets for which the Killing vectors obtained as
bilinears from the Killing spinors are all null.
This analysis completes
the work initiated in \cite{sabragut2007}, where half-supersymmetric
solutions with at least one timelike Killing vector were systematically
classified.
We have also shown that the integrability constraints imposed
by the Killing spinor equations, together with the Bianchi identity
for the 2-form field strengths, are sufficient to imply that the
Einstein, gauge and scalar equations hold automatically.

{}Four classes of solutions were obtained from this analysis:

\begin{itemize}

\item[(i)] In the case where $(\lambda^1_- \neq 0, \lambda^{\bo}_- \neq
0, c_1 \neq 0)$ the metric is given by

\begin{equation} ds^2 = f ds^2(AdS_2) -
(\lambda^{\bo}_-)^2dx^2 - {1\over2\sqrt{2}f}dz^2 -
{f\over2\sqrt{2}}(dy + \beta)^2 \ , \end{equation}

where $ds^2(AdS_2)$ has Ricci scalar $R_{AdS_2} = 4\sqrt{2}c_1^2 $.
$\beta$ is a one form on $AdS_2$ with

\begin{equation}  d\beta =
-2\sqrt{2}c_1 \ {\rm dvol }(AdS_2) \ . \end{equation}

Here $c_1$ is a non-zero constant, and $\lambda^1_-, \lambda^{\bo}_-
\in \bR$. We also find that $f$, $\lambda^1_-, \lambda^{\bo}_-$ and
the scalars $X^I$ are functions of $z$ constrained by

\begin{equation}
\label{finaleqx1}
X_I = {1\over c_1}({K_I \over f} + \chi V_I \lambda^1_- )  \ ,
\end{equation}

\begin{equation}
\label{finaleqx2}
f = {((\lambda^{\bo}_-)^2 +
(\lambda^1_-)^2) \over \sqrt{2}} \ ,
\end{equation}

\begin{equation}
\label{finaleqx3}
\partial_z (f\lambda^1_-) = c_1\lambda^{\bo}_-
 \ ,
 \end{equation}

{}for constant $K_I$. It does not appear to be possible to de-couple
these equations in general.
The field strengths $F^I$ satisfy

\begin{equation} F^I = d(X^I\lambda^1_-  dx) \ . \end{equation}

We remark that although it would appear that these solutions depend on
a free parameter $c_1$, we can without loss of generality set $c_1=1$.
This can be achieved by making the re-scalings
\bea
\lambda^1_- &=& c_1 (\lambda^1_-)', \quad \lambda^{\bar{1}}_- = c_1 (\lambda^{\bar{1}}_-)',
\quad f = c_1^2 f', \quad K_I = c_1^3 (K_I)'
\nn
z &=& c_1 z', \quad x={1 \over c_1} x', \quad y={1 \over c_1} y', \quad \beta = {1 \over c_1} \beta'
\eea
and defining the conformally re-scaled $AdS_2$ factor by
\be
ds^2(AdS_2') = c_1^2 ds^2(AdS_2)
\ee
so that $R_{AdS_2'} = 4 \sqrt{2}$, and $d \beta' = -2\sqrt{2} \ {\rm dvol }(AdS_2')$.
On dropping the primes, it is clear that one can set $c_1=1$ without loss of generality.

\item[(ii)] In the case that $(\lambda^1_- \neq 0, \lambda^{\bo}_- \neq 0, c_1 =
0)$ we find for the metric

\begin{equation} ds^2 = f ds^2(\bR^{1,1}) -
\big( \sqrt{2} f - {c_4^2 \over f^2} \big)dx^2 - {1\over 2\sqrt{2}f}dz^2 -
{f\over 2\sqrt{2}}dy^2 \ , \end{equation}

for non-zero constant $c_4$.
Here the function $f$ and the scalars $X^I$ are constrained by

\begin{equation}
\label{finaleqx4}
\partial_z (fX_I) = \chi V_I (\sqrt{2}f - {c_4^2\over
f^2})^{1\over2} \ , \end{equation}

and the field strengths $F^I$ are given by

\begin{equation} F^I = c_4 \ d \big({X^I \over f}  dx \big) \ . \end{equation}

\item[(iii)] In the case that $(\lambda^1_- = 0, \lambda^{\bo}_- \neq 0)$, we find
that the field strengths vanish, $F^I = 0 $. In addition, the metric
is given by

\begin{equation} ds^2 = h \ ds^2(\bR^{1,3}) - dz^2 \ , \end{equation}

and the scalars satisfy

\begin{equation} \partial_z (h X_I) = 2\chi V_I h  \ . \end{equation}

where $h = (\lambda^{\bo}_-)^2$. This can be seen to be the domain
wall solution found in \cite{sabrklemm2003}, where we identify $h =
(\partial_u f)^{2\over3}$, and $\chi = g$. Note that these solutions can be
obtained from the type $(ii)$ solution described above, by taking the limit $c_4 \rightarrow 0$.

\item[(iv)] In the case that $(\lambda^1_- \neq 0, \lambda^{\bo}_- = 0 )$ we find
that the scalars $X^I$ are constant, and the metric is

\begin{equation} ds^2 = ds^2(AdS_3) - ds^{2}(M_2)  \ . \end{equation}

where $M_2$ is a 2-manifold with Ricci scalar
\be
R_{M_2} = -18 \chi^2 (X^I X^J-Q^{IJ}) V_I V_J
\ee
so $M_2$ is $\mathbb{H}^2$, $\bR^2$, or $S^2$ according as
to whether $(X^I X^J-Q^{IJ}) V_I V_J$ is positive, zero or negative.
Note that in the minimal theory $(X^I X^J-Q^{IJ}) V_I V_J= (V_1)^2$, so
the cases for which $M_2$ is $\bR^2$ or $S^2$ cannot arise in the minimal theory.

The $AdS_3$ manifold has Ricci scalar
\be
R_{AdS_3} = {27 \chi^2 V_I V_J X^I X^J \over 2} \ .
\ee

For the field strengths we find

\begin{equation}  F^I = -3\chi (-X^I V_J X^J + Q^{IJ}V_J) \ {\rm dvol } (M_2)\ . \end{equation}

Note that these product space solutions have previously been found in the context of black
string solutions constructed in \cite{chamssabra99} and \cite{sabraklemm2000c}.

\end{itemize}

\subsection{Interpretation of Solutions}

As we have already stated, the solutions $(iii)$ correspond to domain wall solutions found
in \cite{sabrklemm2003}, and the solutions $(iv)$ correspond to near horizon black string
solutions  \cite{chamssabra99} and \cite{sabraklemm2000c}. We shall therefore concentrate on
solutions $(i)$ and $(ii)$. We shall further assume that
the scalar manifold is symmetric, in which case one has the identity
\be
{9 \over 2} C^{IJK} X_I X_J X_K =1
\ee
where
\be
C^{IJK}= \delta^{II'} \delta^{JJ'} \delta^{KK'} C_{I' J' K'} \ .
\ee

It is then possible to construct the metrics explicitly. We begin with
the solutions of type $(i)$. As mentioned previously, we shall set $c_1=1$ without loss
of generality.To proceed, it is convenient to set
\be
\xi^3 = {9 \over 2} C^{IJK} V_I V_J V_K
\ee
and we assume that $\xi \neq 0$. Also define ${\hat{\rho}}_I$, ${\hat{x}}$ by
\bea
K_I &=& 2 \sqrt{2} C^{-2} {\hat{\rho}}_I
\nn
f \lambda^1_- &=& {2 \sqrt{2}  C^{-2} \over \chi \xi} {\hat{x}}
\eea
for constant $C >0$. Note that ${\hat{x}}$ is not constant. Also set
\bea
{\hat{\alpha}}_0 &=& {9 \over 2} C^{IJK} {\hat{\rho}}_I {\hat{\rho}}_J {\hat{\rho}}_K
\nn
{\hat{\alpha}}_1 &=& {9 \over 2 \xi} C^{IJK} {\hat{\rho}}_I {\hat{\rho}}_J V_K
\nn
{\hat{\alpha}}_2 &=& {9 \over 2 \xi^2} C^{IJK} {\hat{\rho}}_I V_J V_K \ .
\eea
Then ({\ref{finaleqx1}}) implies that
\be
f =2 \sqrt{2} C^{-2} H^{1 \over 3}
\ee
where
\be
H = {\hat{x}}^3 + 3 {\hat{\alpha}}_2 {\hat{x}}^2 + 3 {\hat{\alpha}}_1 {\hat{x}}
+{\hat{\alpha}}_0
\ee
so that the scalars satisfy
\be
H^{1 \over 3} X_I = {\hat{\rho}}_I + {V_I \over \xi} {\hat{x}} \ .
\ee

It is also convenient to introduce co-ordinates $v$, $\rho$ and write the
$AdS_2$ factor in the metric as
\be
ds^2(AdS_2) = -{C^2 \over \sqrt{2}} dv d \rho + { C^4 \over 2 \sqrt{2}} \rho^2 dv^2 \ .
\ee
Finally, on making the re-scalings
\be
x = \chi \xi {\hat{x}}^2, \quad y = C^2 {\hat{x}}^1, \quad \beta = C^2 {\hat{\beta}}
\ee
and using ({\ref{finaleqx3}}), one can rewrite the metric as
\bea
ds^2 &=&H^{1 \over 3} (-2 dv d \rho + C^2 c_1^2 \rho^2 dv^2)
-{4 (\chi \xi)^2 \over C^2} H^{-{2 \over 3}} P (d {\hat{x}}^2)^2
\nn
&-&{H^{1 \over 3} \over 4 (\chi \xi)^2} P^{-1} (d {\hat{x}})^2
-C^2 H^{1 \over 3} (d {\hat{x}}^1 + {\hat{\beta}})^2
\eea
where
\be
\hat{\beta} = \rho dv, \quad P = H - {C^2 \over 4 (\chi \xi)^2} {\hat{x}}^2 \ .
\ee

Finally, define a radial co-ordinate $r$ by
\be
r = H^{1 \over 3} \rho \ .
\ee
It is then straightforward to see that this metric corresponds
to one of the three classes of ``static" local near horizon geometries, written
in Gaussian null co-ordinates, as constructed in \cite{kunduri}
(on dropping the $\hat{}$ on $x, x^1$ and $x^2$).
The horizon is at $r=0$. Note that one can set $C=1$ without
loss of generality, by making appropriately chosen re-scalings, however we retain $C$ here
for ease of comparison. Furthermore, one can also set ${\hat{\alpha}}_2=0$ by making a
constant shift in the ${\hat{x}}$ co-ordinate, this then produces a modification to the function $P$.
It should be noted that a global analysis was carried out in \cite{kunduri} which showed that
the spatial cross-sections of the horizon cannot be regular and compact.

The analysis of the type $(ii)$ solutions is somewhat more straightforward.
In particular, define
\be
\xi^3 = {9 \over 2} C^{IJK} V_I V_J V_K
\ee
and again assume that $\xi \neq 0$, also set
\bea
{{\alpha}}_0 &=& {9 \over 2} C^{IJK} {{\rho}}_I {{\rho}}_J {{\rho}}_K
\nn
{{\alpha}}_1 &=& {9 \over 2 \xi} C^{IJK} {{\rho}}_I {{\rho}}_J V_K
\nn
{{\alpha}}_2 &=& {9 \over 2 \xi^2} C^{IJK} {{\rho}}_I V_J V_K \ .
\eea
It is also convenient to define ${\hat{x}}$ such that
\be
{d \hat{x} \over dz} = \chi \xi \sqrt{ \sqrt{2} f - {c_4^2 \over f^2}}
\ee
so
\be
f X_I = \rho_I + {V_I \over \xi} {\hat{x}}
\ee
and hence
\be
f = \bigg( \hat{x}^3 + 3 \alpha_2 \hat{x}^2 + 3 \alpha_1 \hat{x} + \alpha_0 \bigg)^{1 \over 3} \ .
\ee
Also define ${\hat{x}}^1, {\hat{x}}^2, \tau_0$ by
\bea
x&=& 2^{-{1 \over 4}} {\hat{x}}^2
\nn
y &=& 2^{3 \over 4} C {\hat{x}}^1
\nn
c_4^2 &=& \sqrt{2} \tau_0^3
\eea
for constant $C>0$.
Then the metric can be written as
\bea
ds^2 = 2f dv d \rho - f^{-2} (f^3-\tau_0^3) (d \hat{x}^2)^2
-{f \over 4 (\xi \chi)^2} (f^3- \tau_0^3)^{-1} (d \hat{x})^2 - C^2 f (d \hat{x}^1)^2 \ .
\nn
\eea
On defining the radial co-ordinate $r$ by
\be
r = f^{1 \over 3} \rho
\ee
we recover the second type of ``static" near horizon geometry constructed in
\cite{kunduri}, in  the case for which $\Gamma_0>0$.
The static solutions with $\Gamma_0=0$ found in that paper
correspond to the type $(iii)$ domain wall solutions, with symmetric
scalar manifold. Once more, a global analysis has been constructed,
which shows that these solutions do not correspond
to compact near horizon geometries of regular black holes.
We also remark that it is straightforward to prove that the
solutions of type $(iv)$ correspond to the ``static" solutions
with constant scalars found in \cite{kunduri}.

To summarize, we have shown that
when the scalar manifold is symmetric, and when $C^{IJK} V_I V_J V_K \neq 0$ \footnote{This condition
holds for all solutions of the minimal theory, and also for all asymptotically $AdS_5$ solutions.},
the set of static solutions
found in \cite{kunduri} for which the Killing vector generated from
the Killing spinor is null (excluding the trivial case of the maximally supersymmetric solution $AdS_5$)
is identical to the set of all
half supersymmetric solutions for which all of the Killing spinors
generate null Killing vectors.

\setcounter{section}{0}

\appendix{Integrability Conditions}

The gravitino and dilatino integrability conditions, respectively,
can be put into the form

\begin{equation} \label{EGS1} (E_{\alpha}{}^{\beta} \Gamma_{\beta} + {1\over3}G^{\beta}\Gamma_{\alpha\beta}
-{2\over3}G_{\alpha})\epsilon = 0 \ , \end{equation}

\begin{equation} \label{EGS2} (S_I - {2\over3}(G_{I\alpha} -
X_I X^J G_{J\alpha})\Gamma^{\alpha})\epsilon = 0 \ , \end{equation}

acting on a Dirac spinor $\epsilon = \lambda^1_+\psi^1_+ +
\lambda^{\bo}_+\psi^{\bo}_+ + \lambda^1_-\psi^1_- +
\lambda^{\bo}_-\psi^{\bo}_-$. Here

\bea E_{\alpha\beta} &=& R_{\alpha\beta} +
Q_{IJ}F^I{}_{\alpha\mu}F^J{}_{\beta}{}^{\mu} - Q_{IJ}\nabla_{\alpha}
X^I\nabla_{\beta} X^J \nn &+&
g_{\alpha\beta}\big(-{1\over6}Q_{IJ}F^I_{\beta_1
\beta_2}F^{J\beta_1\beta_2} +
6\chi^2({1\over2}Q^{IJ}-X^IX^J)V_IV_J\big) \ , \eea

\begin{equation} G_{I\alpha} = \nabla^{\beta}(Q_{IJ}F^J{}_{\alpha\beta}) +
{1\over16}C_{IJK}\epsilon_{\alpha}{}^{\beta_1\beta_2\beta_3\beta_4}F^J{}_{\beta_1\beta_2}F^K{}_{\beta_3\beta_4}
\ , \end{equation}

\bea S_I &=& \nabla^{\alpha}\nabla_{\alpha} X_I - ({1\over6}C_{MNI}
- {1\over2}X_IC_{MNJ}X^J)\nabla_{\alpha} X^M \nabla^{\alpha} X^N \nn
&-&{1\over2}\big(X_MX^PC_{NPI}  - {1\over6}C_{MNI} - 6X_IX_MX_N +
{1\over6}X_IC_{MNJ}X^J
\big)F^M{}_{\beta_1\beta_2}F^{N\beta_1\beta_2} \nn
&-&3\chi^2V_MV_N\big({1\over2}Q^{ML}Q^{NP}C_{LPI} + X_I(Q^{MN}
-2X^MX^N)\big) \ , \eea

with $G_{\beta} = X^IG_{I\beta}$. The $\psi^1_+, \psi^{\bo}_+,
\psi^1_-, \psi^{\bo}_-$ components of ({\ref{EGS1}}) are
respectively, for $ \alpha = + $

\begin{equation} \label{EGS1+1} \sqrt{2}E_{+-}\lambda^1_- +
\sqrt{2}iE_{+1}\lambda^{\bo}_+ - iE_{+2}\lambda^1_+ +
{1\over3}(-G_+\lambda^1_+ - 2iG_1 \lambda^{\bo}_- +
\sqrt{2}iG_2\lambda^1_-) = 0 \ ,
\end{equation}

\begin{equation} \label{EGS1+2} \sqrt{2}E_{+-}\lambda^{\bo}_- +
\sqrt{2}iE_{+\bo}\lambda^{1}_+ - iE_{+2}\lambda^{\bo}_+ -
{1\over3}(G_+\lambda^{\bo}_+ + 2iG_{\bo}\lambda^{1}_- +
\sqrt{2}iG_2\lambda^{\bo}_-)  = 0 \ ,
\end{equation}

\begin{equation} \label{EGS1+3} \sqrt{2} E_{++}\lambda^1_+ -
\sqrt{2}iE_{+1}\lambda^{\bo}_- + iE_{+2} -\lambda^1_- - G_+
\lambda^1_-  = 0 \ , \end{equation}

\begin{equation} \label{EGS1+4} \sqrt{2} E_{++}\lambda^{\bo}_+ -
\sqrt{2}iE_{+\bo}\lambda^1_- - iE_{+2} -\lambda^{\bo}_- - G_+
\lambda^{\bo}_-  = 0 \ .\end{equation}

For $ \alpha= - $

\begin{equation} \label{EGS1-1} \sqrt{2}E_{--}\lambda^1_- +
\sqrt{2}iE_{-1}\lambda^{\bo}_+ -iE_{-2}\lambda^1_+ - G_-\lambda^1_+
 = 0 \ ,
\end{equation}

\begin{equation} \label{EGS1-2} \sqrt{2}E_{--}\lambda^{\bo}_- +
\sqrt{2}iE_{-\bo}\lambda^1_+ -iE_{-2}\lambda^{\bo}_+ -
G_-\lambda^{\bo}_+ = 0 \ ,
\end{equation}

\begin{equation} \label{EGS1-3} \sqrt{2}E_{-+}\lambda^1_+ -
\sqrt{2}iE_{-1}\lambda^{\bo}_- +iE_{-2}\lambda^1_- +
{1\over3}(-G_-\lambda^1_- + 2iG_1\lambda^{\bo}_+
-\sqrt{2}iG_2\lambda^1_+)  = 0 \ ,
\end{equation}

\begin{equation} \label{EGS1-4} \sqrt{2}E_{-+}\lambda^{\bo}_+ -
\sqrt{2}iE_{-\bo}\lambda^1_- +iE_{-2}\lambda^{\bo}_- +
{1\over3}(-G_-\lambda^{\bo}_- + 2iG_{\bo}\lambda^1_+
+\sqrt{2}iG_2\lambda^{\bo}_+) = 0 \ .
\end{equation}

For $\alpha = 1$

\begin{equation} \label{EGS111} \sqrt{2}E_{1-}\lambda^1_- +
\sqrt{2}iE_{11}\lambda^{\bo}_+ - iE_{12}\lambda^1_+ - G_1\lambda^1_+
 = 0 \ ,
\end{equation}

\begin{equation} \label{EGS112} \sqrt{2}E_{1-}\lambda^{\bo}_- +
\sqrt{2}iE_{1\bo}\lambda^1_+ + iE_{12}\lambda^{\bo}_+ +
{1\over3}(-2iG_-\lambda^1_- -\sqrt{2}G_2\lambda^1_+ -
G_1\lambda^{\bo}_+)  = 0 \ ,
\end{equation}

\begin{equation} \label{EGS113} \sqrt{2}E_{1+}\lambda^1_+ -
\sqrt{2}iE_{11}\lambda^{\bo}_- + iE_{12}\lambda^1_- - G_1\lambda^1_-
= 0 \ ,
\end{equation}

\begin{equation} \label{EGS114} \sqrt{2}E_{1+}\lambda^{\bo}_+ -
\sqrt{2}iE_{1\bo}\lambda^1_- - iE_{12}\lambda^{\bo}_- +
{1\over3}(2iG_+\lambda^1_+ - \sqrt{2}G_2\lambda^1_- -
G_1\lambda^{\bo}_-)  = 0 \ .
\end{equation}

For $\alpha = \bo $

\begin{equation} \label{EGS1bar1} \sqrt{2}E_{\bo-}\lambda^1_- +
\sqrt{2}iE_{\bo1}\lambda^{\bo}_+ - iE_{\bo2}\lambda^1_+ +
{1\over3}(-2iG_-\lambda^{\bo}_- +\sqrt{2}G_2\lambda^{\bo}_+ -
G_{\bo}\lambda^1_+)  = 0 \ ,
\end{equation}

\begin{equation} \label{EGS1bar2} \sqrt{2}E_{\bo-}\lambda^{\bo}_- +
\sqrt{2}iE_{\bo\bo}\lambda^1_+ + iE_{\bo2}\lambda^{\bo}_+ -
G_{\bo}\lambda^{\bo}_+  = 0 \ ,
\end{equation}

\begin{equation} \label{EGS1bar3} \sqrt{2}E_{\bo+}\lambda^1_+ -
\sqrt{2}iE_{\bo1}\lambda^{\bo}_- + iE_{\bo2}\lambda^1_- +
{1\over3}(2iG_+\lambda^{\bo}_+ + \sqrt{2}G_2\lambda^{\bo}_- -
G_{\bo}\lambda^1_-)  = 0 \ ,
\end{equation}

\begin{equation} \label{EGS1bar4} \sqrt{2}E_{\bo+}\lambda^{\bo}_+ -
\sqrt{2}iE_{\bo\bo}\lambda^1_- - iE_{\bo2}\lambda^{\bo}_- -
G_{\bo}\lambda^{\bo}_-  = 0 \ .
\end{equation}

Finally for $\alpha = 2 $ we have

\begin{equation} \label{EGS121} \sqrt{2}E_{2-}\lambda^1_- +
\sqrt{2}iE_{21}\lambda^{\bo}_+ - iE_{22}\lambda^1_+ +
{1\over3}(\sqrt{2}iG_-\lambda^1_- - \sqrt{2}G_1\lambda^{\bo}_+) -
{2\over3}G_2\lambda^1_+ = 0 \ , \end{equation}

\begin{equation} \label{EGS122} \sqrt{2}E_{2-}\lambda^{\bo}_- +
\sqrt{2}iE_{2\bo}\lambda^1_+ + iE_{22}\lambda^{\bo}_+ +
{1\over3}(-\sqrt{2}iG_-\lambda^{\bo}_- +\sqrt{2}G_{\bo}\lambda^1_+)
-  {2\over3}G_2\lambda^{\bo}_+ = 0 \ , \end{equation}

\begin{equation} \label{EGS123} \sqrt{2}E_{2+}\lambda^1_+ -
\sqrt{2}iE_{21}\lambda^{\bo}_- + iE_{22}\lambda^1_- +
{1\over3}(\sqrt{2}iG_+\lambda^1_+ - \sqrt{2}G_1\lambda^{\bo}_- ) -
{2\over3}G_2\lambda^1_- = 0 \ , \end{equation}

\begin{equation} \label{EGS124} \sqrt{2}E_{2+}\lambda^{\bo}_+ -
\sqrt{2}iE_{2\bo}\lambda^1_- - iE_{22}\lambda^{\bo}_- +
{1\over3}(-\sqrt{2}iG_+\lambda^{\bo}_+ + \sqrt{2}G_{\bo}\lambda^1_-)
-  {2\over3}G_2\lambda^{\bo}_- = 0 \ . \end{equation}

Acting on the first Killing spinor $ \epsilon = \psi^1_+ $, we find
the following constraints

\begin{equation}  E_{++} = E_{+2} = E_{+1} = E_{-+} = E_{-1} = E_{-2}  = 0 \ ,  \end{equation}

and

\begin{equation} E_{1+} = E_{11} = E_{12} = E_{1\bo}  = 0 \ ,  \end{equation}

as well as

\begin{equation} E_{2+} = E_{21}= E_{22} = 0 \ , \end{equation}

together with

\begin{equation} G_+ = G_- = G_2 = G_1 = 0 \ . \end{equation}

We can then substitute these back, finding the following
non-vanishing constraints for $\alpha = +$

\begin{equation} E_{+-}\lambda^{\bo}_- = E_{+-}\lambda^1_- = 0 \ , \end{equation}

for $\alpha = - $

\begin{equation} E_{--} \lambda^1_- = E_{--} \lambda^{\bo}_- = 0  \ , \end{equation}

for $\alpha = 1$

\begin{equation} E_{1-} \lambda^1_- = E_{1-} \lambda^{\bo}_- = 0  \ , \end{equation}

for $\alpha = 2$

\begin{equation} E_{2-}\lambda^1_- = E_{2-} \lambda^{\bo}_- = 0 \ .\end{equation}

We recall from Section 4 that the residual gauge transformations
preserving $\epsilon = \psi^1_+$ allowed us to place our second
Killing spinor $\eta = \lambda^1_+ \psi^1_+ + \lambda^{\bo}_+
\psi^{\bo}_+ + \lambda^1_- \psi^1_- + \lambda^{\bo}_- \psi^{\bo}_- $
into a form where either $\lambda^{\alpha}_- = 0 $, or
$\lambda^{\alpha}_+ = 0 $, for $\alpha = 1, \bo$. In section 6
solutions with $\lambda^{\alpha}_- = 0$ were found to be only
${1\over4}$ supersymmetric. If we then examine the case
$\lambda^{\alpha}_+ = 0 $, we see that we must have $G = X^IG_I = 0$
and $E = 0$.

Evaluating ({\ref{EGS2}}) for a general Dirac spinor $\epsilon$,
yields, for the $\psi^1_+, \psi^{\bo}_+, \psi^1_-, \psi^{\bo}_-$
components

\begin{equation} \label{EGS21} S_I\lambda^1_+ -
{2\over3}(\sqrt{2}G_{I-}\lambda^1_- + \sqrt{2}iG_{I1}\lambda^{\bo}_+
- iG_{I2}\lambda^1_+) = 0 \ , \end{equation}

\begin{equation} \label{EGS22} S_I\lambda^{\bo}_+ -
{2\over3}(\sqrt{2}G_{I-}\lambda^{\bo}_- +
\sqrt{2}iG_{I\bo}\lambda^1_+ + iG_{I2}\lambda^{\bo}_+) = 0 \ ,
\end{equation}

\begin{equation}\label{EGS23} S_I\lambda^1_- -
{2\over3}(\sqrt{2}G_{I+}\lambda^1_+ - \sqrt{2}iG_{I1}\lambda^{\bo}_-
+ iG_{I2}\lambda^1_-) = 0 \ , \end{equation}

\begin{equation} \label{EGS24} S_I\lambda^{\bo}_- -
{2\over3}(\sqrt{2}G_{I+}\lambda^{\bo}_+
-\sqrt{2}iG_{I\bo}\lambda^1_- - iG_{I2}\lambda^{\bo}_-) = 0 \ ,
\end{equation}

where we have used $G = X^IG_I = 0$. Next, we restrict to the case
$\epsilon = \psi^1_+$

\begin{equation} S_I = 0 \ , \end{equation}

\begin{equation} G_{I2} = 0 \ , \end{equation}

\begin{equation} G_{I\bo} = 0  \ , \end{equation}

\begin{equation} G_{I+} = 0 \ . \end{equation}

Substituting back, we find that

\begin{equation} G_{I-}\lambda^1_- = G_{I-}\lambda^{\bo}_- = 0 \ , \end{equation}

so $G_I = 0$ and $S_I$ = 0.

\appendix{The Linear System}

In this appendix we present the decomposition of the Killing spinor
equations acting on a generic Killing spinor (written in an adapted
null basis), and then present a special case.

\setcounter{subsection}{0}

\subsection{Solutions with $ \epsilon = \lambda^\alpha_+ \psi^\alpha_+ + \lambda^\alpha_- \psi^\alpha_- $ }

The action of the dilatino equations on $ \epsilon $
is:

\begin{eqnarray} 4i\chi(X^IV_JX^J - {3\over2}Q^{IJ}V_J)\lambda_+^1 +
2\sqrt{2}\partial_-X^I \lambda_-^1 + 2\sqrt{2}i\partial_{1}X^I
\lambda_+^{\bar{1}} - 2i\partial_2X^I \lambda_+^1 \nn + 2(F^I_{+-} -
X^IH_{+-})\lambda_+^1 +4i(F^{I}_{-1} -
X^IH_{-1})\lambda_-^{\bar{1}}-2\sqrt{2}i(F^{I}_{-2} -
X^IH_{-2})\lambda_-^1   \nn +  2\sqrt{2}(F^{I}_{12} -
X^IH_{12})\lambda_+^{\bar{1}} +2(F^{I}_{1\bar{1}} -
X^IH_{1\bar{1}})\lambda_+^1 = 0 \ , \end{eqnarray}

\begin{eqnarray} 4i\chi(X^IV_JX^J - {3\over2}Q^{IJ}V_J)\lambda_+^{\bar{1}} +
2\sqrt{2}\partial_-X^I \lambda_-^{\bar{1}} +
2\sqrt{2}i\partial_{\bo}X^I \lambda_+^1 + 2i\partial_2X^I
\lambda_+^{\bar{1}}  \nn + 2(F^{I}_{+-} -
X^IH_{+-})\lambda_+^{\bar{1}} + 4i(F^{I}_{-\bo} -
X^IH_{-\bo})\lambda_-^1 + 2\sqrt{2}i(F^{I}_{-2} -
X^IH_{-2})\lambda_-^{\bar{1}}  \nn - 2\sqrt{2}(F^{I}_{\bo2} -
X^IH_{\bo2})\lambda_+^1 -2(F^{I}_{1\bar{1}} -
X^IH_{1\bar{1}})\lambda_+^{\bar{1}} = 0 \ , \end{eqnarray}

\begin{eqnarray} 4i\chi(X^IV_JX^J - {3\over2}Q^{IJ}V_J)\lambda_-^1 +
2\sqrt{2}\partial_+X^I \lambda_+^1 - 2\sqrt{2}i\partial_1X^I
\lambda_-^{\bar{1}} + 2i\partial_2X^I \lambda_-^1 \nn - 2(F^{I}_{+-}
- X^IH_{+-})\lambda_-^1 -4i(F^{I}_{+1} -
X^IH_{+1})\lambda_+^{\bar{1}}+2\sqrt{2}i(F^{I}_{+2} -
X^IH_{+2})\lambda_+^1  \nn +  2\sqrt{2}(F^{I}_{12} -
X^IH_{12})\lambda_-^{\bar{1}} +2(F^{I}_{1\bar{1}} -
X^IH_{1\bar{1}})\lambda_-^1 = 0 \ , \end{eqnarray}

\begin{eqnarray} 4i\chi(X^IV_JX^J - {3\over2}Q^{IJ}V_J)\lambda_-^{\bar{1}} +
2\sqrt{2}\partial_+X^I \lambda_+^{\bar{1}} -
2\sqrt{2}i\partial_{\bo}X^I \lambda_-^1 - 2i\partial_2X^I
\lambda_-^{\bar{1}} \nn - 2(F^{I}_{+-} -
X^IH_{+-})\lambda_-^{\bar{1}} -4i(F^{I}_{+\bo} -
X^IH_{+\bo})\lambda_+^1 - 2\sqrt{2}i(F^{I}_{+2} -
X^IH_{+2})\lambda_+^{\bar{1}}  \nn - 2\sqrt{2}(F^{I}_{\bo2} -
X^IH_{\bo2})\lambda_-^1 -2(F^{I}_{1\bar{1}} -
X^IH_{1\bar{1}})\lambda_-^{\bar{1}} = 0 \ . \end{eqnarray}

The action of the gravitino equation on $\epsilon$ in the $+$ direction is given by
(taking the $\psi_+^1 ,
\psi_+^{\bar{1}}, \psi_-^1, \psi_-^{\bar{1}}$ components in turn):

\begin{eqnarray} (\partial_+ - {3i\chi\over2}A_+)\lambda_+^1
-{3\over4}(\sqrt{2}H_{+-}\lambda^1_-
+\sqrt{2}iH_{+1}\lambda^{\bar{1}}_+ - iH_{+2}\lambda_+^1) \nn +
{1\over2}(-\omega_{+,+-}\lambda^1_+ -
2i\omega_{+,-1}\lambda^{\bar{1}}_-
+\sqrt{2}i\omega_{+,-2}\lambda^1_-
-\sqrt{2}\omega_{+,12}\lambda_+^{\bar{1}}
-\omega_{+,1\bar{1}}\lambda_+^1) \nn +
{\sqrt{2}\over4}(H_{+-}\lambda^1_- +2iH_{+1}\lambda^{\bar{1}}_+ -
\sqrt{2}iH_{+2}\lambda^1_+ -\sqrt{2}H_{12}\lambda^{\bar{1}}_- -
H_{1\bar{1}}\lambda^1_-) \nn + {i\chi\over\sqrt{2}}V_IX^I\lambda^1_-
= 0 \ , \end{eqnarray}

\begin{eqnarray} (\partial_+ - {3i\chi\over2}A_+)\lambda_+^{\bar{1}}
-{3\over4}(\sqrt{2}H_{+-}\lambda^{\bar{1}}_-
+\sqrt{2}iH_{+\bo}\lambda^{1}_+ + iH_{+2}\lambda_+^{\bar{1}} )\nn+
{1\over2}(-\omega_{+,+-}\lambda^{\bar{1}}_+ -
2i\omega_{+,-\bar{1}}\lambda^{1}_-
-\sqrt{2}i\omega_{+,-2}\lambda^{\bar{1}}_-
+\sqrt{2}\omega_{+,\bo2}\lambda_+^{1}
+\omega_{+,1\bar{1}}\lambda_+^{\bar{1}}) \nn +
{\sqrt{2}\over4}(H_{+-}\lambda^{\bar{1}}_- +2iH_{+\bo}\lambda^{1}_+
+ \sqrt{2}iH_{+2}\lambda^{\bar{1}}_+ +\sqrt{2}H_{\bo2}\lambda^1_- +
H_{1\bar{1}}\lambda^{\bar{1}}_-) \nn +
{i\chi\over\sqrt{2}}V_IX^I\lambda^{\bar{1}}_- = 0 \ , \end{eqnarray}

\begin{eqnarray} (\partial_+ - {3i\chi\over2}A_+)\lambda_-^1
-{3\over4}(-\sqrt{2}iH_{+1}\lambda^{\bar{1}}_- +iH_{+2}\lambda_-^1)
\nn + {1\over2}(\omega_{+,+-}\lambda^1_- +
2i\omega_{+,+1}\lambda^{\bar{1}}_+
-\sqrt{2}i\omega_{+,+2}\lambda^1_+
-\sqrt{2}\omega_{+,12}\lambda_-^{\bar{1}} -
\omega_{+,1\bar{1}}\lambda_-^1) = 0 \ , \end{eqnarray}

\begin{eqnarray} (\partial_+ - {3i\chi\over2}A_+)\lambda_-^{\bar{1}}
-{3\over4}(-\sqrt{2}iH_{+\bo}\lambda^1_- -
iH_{+2}\lambda_-^{\bar{1}}) \nn +
{1\over2}(\omega_{+,+-}\lambda^{\bar{1}}_- +
2i\omega_{+,+\bo}\lambda^1_+
+\sqrt{2}i\omega_{+,+2}\lambda^{\bar{1}}_+
+\sqrt{2}\omega_{+,\bo2}\lambda_-^{1}
+\omega_{+,1\bar{1}}\lambda_-^{\bar{1}}) = 0 \ .\end{eqnarray}

In the $-$ direction

\begin{eqnarray} (\partial_- - {3i\chi\over2}A_-)\lambda_+^1 -{3\over4}(
\sqrt{2}iH_{-1}\lambda^{\bar{1}}_+ - iH_{-2}\lambda_+^1) \nn +
{1\over2}(-\omega_{-,+-}\lambda^1_+ -
2i\omega_{-,-1}\lambda^{\bar{1}}_-
+\sqrt{2}i\omega_{-,-2}\lambda^1_-
-\sqrt{2}\omega_{-,12}\lambda_+^{\bar{1}} -
\omega_{-,1\bar{1}}\lambda_+^1) = 0 \ , \end{eqnarray}

\begin{eqnarray} (\partial_- - {3i\chi\over2}A_-)\lambda_+^{\bar{1}}
-{3\over4}( \sqrt{2}iH_{-\bo}\lambda^{1}_+ +
iH_{-2}\lambda_+^{\bar{1}} ) \nn +
{1\over2}(-\omega_{-,+-}\lambda^{\bar{1}}_+ -
2i\omega_{-,-\bo}\lambda^{1}_-
-\sqrt{2}i\omega_{-,-2}\lambda^{\bar{1}}_-
+\sqrt{2}\omega_{-,\bo2}\lambda_+^{1}
+\omega_{-,1\bar{1}}\lambda_+^{\bar{1}}) = 0 \ , \end{eqnarray}

\begin{eqnarray}(\partial_- - {3i\chi\over2}A_-)\lambda_-^1
-{3\over4}(-\sqrt{2}H_{+-}\lambda^1_+
-\sqrt{2}iH_{-1}\lambda^{\bar{1}}_- + iH_{-2}\lambda_-^1) \nn +
{1\over2}(\omega_{-,+-}\lambda^1_- +
2i\omega_{-,+1}\lambda^{\bar{1}}_+
-\sqrt{2}i\omega_{-,+2}\lambda^1_+
-\sqrt{2}\omega_{-,12}\lambda_-^{\bar{1}} -
\omega_{-,1\bar{1}}\lambda_-^1)  \nn +
{\sqrt{2}\over4}(-H_{+-}\lambda^1_+ -2iH_{-1}\lambda^{\bar{1}}_- +
\sqrt{2}iH_{-2}\lambda^1_- -\sqrt{2}H_{12}\lambda^{\bar{1}}_+ -
H_{1\bar{1}}\lambda^1_+) \nn + {i\chi\over\sqrt{2}}V_IX^I\lambda^1_+
= 0 \ , \end{eqnarray}

\begin{eqnarray} (\partial_- - {3i\chi\over2}A_-)\lambda_-^{\bar{1}}
-{3\over4}(-\sqrt{2}H_{+-}\lambda^{\bar{1}}_+
-\sqrt{2}iH_{-\bo}\lambda^1_- - iH_{-2}\lambda_-^{\bar{1}}) \nn +
{1\over2}(\omega_{-,+-}\lambda^{\bar{1}}_- +
2i\omega_{-,+\bo}\lambda^1_+
+\sqrt{2}i\omega_{-,+2}\lambda^{\bar{1}}_+
+\sqrt{2}\omega_{-,\bo2}\lambda_-^{1}
+\omega_{-,1\bar{1}}\lambda_-^{\bar{1}})  \nn +
{\sqrt{2}\over4}(-H_{+-}\lambda^{\bar{1}}_+ -2iH_{-\bo}\lambda^{1}_-
- \sqrt{2}iH_{-2}\lambda^{\bar{1}}_- +\sqrt{2}H_{\bo2}\lambda^1_+ +
H_{1\bar{1}}\lambda^{\bar{1}}_+) \nn +
{i\chi\over\sqrt{2}}V_IX^I\lambda^{\bar{1}}_+= 0 \ . \end{eqnarray}

In the $1$ direction

\begin{eqnarray} (\partial_1 - {3i\chi\over2}A_1)\lambda_+^1
-{3\over4}(-\sqrt{2}H_{-1}\lambda^1_- - iH_{12}\lambda_+^1) \nn +
{1\over2}(-\omega_{1,+-}\lambda^1_+ -
2i\omega_{1,-1}\lambda^{\bar{1}}_-
+\sqrt{2}i\omega_{1,-2}\lambda^1_-
-\sqrt{2}\omega_{1,12}\lambda_+^{\bar{1}} -
\omega_{1,1\bar{1}}\lambda_+^1) = 0 \ , \end{eqnarray}

\begin{eqnarray} (\partial_1 - {3i\chi\over2}A_1)\lambda_+^{\bar{1}}
-{3\over4}(-\sqrt{2}H_{-1}\lambda^{\bar{1}}_-
+\sqrt{2}iH_{1\bo}\lambda^{1}_+ + iH_{12}\lambda_+^{\bar{1}}) \nn +
{1\over2}(-\omega_{1,+-}\lambda^{\bar{1}}_+ -
2i\omega_{1,-\bo}\lambda^{1}_-
-\sqrt{2}i\omega_{1,-2}\lambda^{\bar{1}}_-
+\sqrt{2}\omega_{1,\bo2}\lambda_+^{1}
+\omega_{1,1\bar{1}}\lambda_+^{\bar{1}})\nn
-{i\sqrt{2}\over4}(-H_{+-}\lambda^1_+ -2iH_{-1}\lambda^{\bar{1}}_- +
\sqrt{2}iH_{-2}\lambda^1_- -\sqrt{2}H_{12}\lambda^{\bar{1}}_+ -
H_{1\bar{1}}\lambda^1_+) \nn + {\chi\over\sqrt{2}}V_IX^I\lambda^1_+
= 0 \ , \end{eqnarray}

\begin{eqnarray} (\partial_1 - {3i\chi\over2}A_1)\lambda_-^1
-{3\over4}(-\sqrt{2}H_{+1}\lambda^1_+ + iH_{12}\lambda_-^1) \nn +
{1\over2}(\omega_{1,+-}\lambda^1_- +
2i\omega_{1,+1}\lambda^{\bar{1}}_+
-\sqrt{2}i\omega_{1,+2}\lambda^1_+
-\sqrt{2}\omega_{1,12}\lambda_-^{\bar{1}} -
\omega_{1,1\bar{1}}\lambda_-^1) = 0 \ , \end{eqnarray}

\begin{eqnarray} (\partial_1 - {3i\chi\over2}A_1)\lambda_-^{\bar{1}}
-{3\over4}(-\sqrt{2}H_{+1}\lambda^{\bar{1}}_+
-\sqrt{2}iH_{1\bo}\lambda^1_- + iH_1{}^2\lambda_-^{\bar{1}}) \nn +
{1\over2}(\omega_{1,+-}\lambda^{\bar{1}}_- +
2i\omega_{1,+\bo}\lambda^1_+
+\sqrt{2}i\omega_{1,+2}\lambda^{\bar{1}}_+
+\sqrt{2}\omega_{1,\bo2}\lambda_-^{1}
+\omega_{1,1\bar{1}}\lambda_-^{\bar{1}})  \nn +
{\sqrt{2}i\over4}(H_{+-}\lambda^1_- +2iH_{+1}\lambda^{\bar{1}}_+ -
\sqrt{2}iH_{+2}\lambda^1_+ -\sqrt{2}H_{12}\lambda^{\bar{1}}_- -
H_{1\bar{1}}\lambda^1_-) \nn -  {\chi\over\sqrt{2}}V_IX^I\lambda^1_-
= 0 \ . \end{eqnarray}

In the $\bar{1}$ direction

\begin{eqnarray} (\partial_{\bar{1}} - {3i\chi\over2}A_{\bar{1}})\lambda_+^1
-{3\over4}(-\sqrt{2}H_{-\bar{1}}\lambda^1_-
-\sqrt{2}iH_{1\bar{1}}\lambda^{\bar{1}}_+ -
iH_{\bar{1}2}\lambda_+^1) \nn +
{1\over2}(-\omega_{\bar{1},+-}\lambda^1_+ -
2i\omega_{\bo,-1}\lambda^{\bar{1}}_-
+\sqrt{2}i\omega_{\bo,-2}\lambda^1_-
-\sqrt{2}\omega_{\bar{1},12}\lambda_+^{\bar{1}} -
\omega_{\bo,1\bar{1}}\lambda_+^1) \nn
-{\sqrt{2}i\over4}(-H_{+-}\lambda^{\bar{1}}_+
-2iH_{-\bo}\lambda^{1}_- - \sqrt{2}iH_{-2}\lambda^{\bar{1}}_-
+\sqrt{2}H_{\bo2}\lambda^1_+ + H_{1\bar{1}}\lambda^{\bar{1}}_+) \nn
+ {\chi\over\sqrt{2}}V_IX^I\lambda^{\bar{1}}_+ = 0 \ ,
\end{eqnarray}

\begin{eqnarray} (\partial_{\bar{1}} -
{3i\chi\over2}A_{\bar{1}})\lambda_+^{\bar{1}}
-{3\over4}(-\sqrt{2}H_{-\bar{1}}\lambda^{\bar{1}}_- +
iH_{\bar{1}2}\lambda_+^{\bar{1}}) \nn +
{1\over2}(-\omega_{\bo,+-}\lambda^{\bar{1}}_+ -
2i\omega_{\bo,-\bo}\lambda^{1}_-
-\sqrt{2}i\omega_{\bo,-2}\lambda^{\bar{1}}_-
+\sqrt{2}\omega_{\bo,\bo2}\lambda_+^{1}
+\omega_{\bo,1\bar{1}}\lambda_+^{\bar{1}}) = 0 \ , \end{eqnarray}

\begin{eqnarray} (\partial_{\bar{1}} - {3i\chi\over2}A_{\bar{1}})\lambda_-^1
-{3\over4}(-\sqrt{2}H_{+\bar{1}}\lambda^1_+
+\sqrt{2}iH_{1\bar{1}}\lambda^{\bar{1}}_- +
iH_{\bar{1}2}\lambda_-^1) \nn + {1\over2}(\omega_{\bo,+-}\lambda^1_-
+ 2i\omega_{\bo,+1}\lambda^{\bar{1}}_+
-\sqrt{2}i\omega_{\bo,+2}\lambda^1_+
-\sqrt{2}\omega_{\bo,12}\lambda_-^{\bar{1}} -
\omega_{\bo,1\bar{1}}\lambda_-^1)  \nn +
{\sqrt{2}i\over4}(H_{+-}\lambda^{\bar{1}}_- +2iH_{+\bo}\lambda^{1}_+
+ \sqrt{2}iH_{+2}\lambda^{\bar{1}}_+ +\sqrt{2}H_{\bo2}\lambda^1_- +
H_{1\bar{1}}\lambda^{\bar{1}}_-) \nn -
{\chi\over\sqrt{2}}V_IX^I\lambda^{\bar{1}}_- = 0 \ , \end{eqnarray}

\begin{eqnarray} (\partial_{\bar{1}} -
{3i\chi\over2}A_{\bar{1}})\lambda_-^{\bar{1}}
-{3\over4}(-\sqrt{2}H_{+\bar{1}}\lambda^{\bar{1}}_+ -
iH_{\bar{1}2}\lambda_-^{\bar{1}}) \nn +
{1\over2}(\omega_{\bo,+-}\lambda^{\bar{1}}_- +
2i\omega_{\bo,+\bo}\lambda^1_+
+\sqrt{2}i\omega_{\bo,+2}\lambda^{\bar{1}}_+
+\sqrt{2}\omega_{\bo,\bo2}\lambda_-^{1}
+\omega_{\bo,1\bar{1}}\lambda_-^{\bar{1}}) = 0 \ . \end{eqnarray}

Finally, in the $2$ direction

\begin{eqnarray} (\partial_2 - {3i\chi\over2}A_2)\lambda_+^1
-{3\over4}(-\sqrt{2}H_{-2}\lambda^1_-
-\sqrt{2}iH_{12}\lambda^{\bar{1}}_+) \nn +
{1\over2}(-\omega_{2,+-}\lambda^1_+ -
2i\omega_{2,-1}\lambda^{\bar{1}}_-
+\sqrt{2}i\omega_{2,-2}\lambda^1_-
-\sqrt{2}\omega_{2,12}\lambda_+^{\bar{1}} -
\omega_{2,1\bar{1}}\lambda_+^1) \nn + {i\over4}(-H_{+-}\lambda^1_+
-2iH_{-1}\lambda^{\bar{1}}_- + \sqrt{2}iH_{-2}\lambda^1_-
-\sqrt{2}H_{12}\lambda^{\bar{1}}_+ - H_{1\bar{1}}\lambda^1_+) \nn -
{\chi\over2}V_IX^I\lambda^1_+ = 0 \ , \end{eqnarray}

\begin{eqnarray} (\partial_2 - {3i\chi\over2}A_2)\lambda_+^{\bar{1}}
-{3\over4}(-\sqrt{2}H_{-2}\lambda^{\bar{1}}_-
-\sqrt{2}iH_{\bo2}\lambda^{1}_+) \nn +
{1\over2}(-\omega_{2,+-}\lambda^{\bar{1}}_+ -
2i\omega_{2,-\bo}\lambda^{1}_-
-\sqrt{2}i\omega_{2,-2}\lambda^{\bar{1}}_-
+\sqrt{2}\omega_{2,\bo2}\lambda_+^{1}
+\omega_{2,1\bar{1}}\lambda_+^{\bar{1}}) \nn -
{i\over4}(-H_{+-}\lambda^{\bar{1}}_+ -2iH_{-\bo}\lambda^{1}_- -
\sqrt{2}iH_{-2}\lambda^{\bar{1}}_- +\sqrt{2}H_{\bo2}\lambda^1_+ +
H_{1\bar{1}}\lambda^{\bar{1}}_+) \nn +
{\chi\over2}V_IX^I\lambda^{\bar{1}}_+ = 0 \ , \end{eqnarray}

\begin{eqnarray} (\partial_2 - {3i\chi\over2}A_2)\lambda_-^1
-{3\over4}(-\sqrt{2}H_{+2}\lambda^1_+
+\sqrt{2}iH_{12}\lambda^{\bar{1}}_-) \nn +
{1\over2}(\omega_{2,+-}\lambda^1_- +
2i\omega_{2,+1}\lambda^{\bar{1}}_+
-\sqrt{2}i\omega_{2,+2}\lambda^1_+
-\sqrt{2}\omega_{2,12}\lambda_-^{\bar{1}} -
\omega_{2,1\bar{1}}\lambda_-^1) \nn - {i\over4}(H_{+-}\lambda^1_-
+2iH_{+1}\lambda^{\bar{1}}_+ - \sqrt{2}iH_{+2}\lambda^1_+
-\sqrt{2}H_{12}\lambda^{\bar{1}}_- - H_{1\bar{1}}\lambda^1_-) +
{\chi\over2}V_IX^I\lambda^1_- = 0 \ , \nn \end{eqnarray}

\begin{eqnarray} (\partial_2 - {3i\chi\over2}A_2)\lambda_-^{\bar{1}}
-{3\over4}(-\sqrt{2}H_{+2}\lambda^{\bar{1}}_+
+\sqrt{2}iH_{\bo2}\lambda^1_-) \nn +
{1\over2}(\omega_{2,+-}\lambda^{\bar{1}}_- +
2i\omega_{2,+\bo}\lambda^1_+
+\sqrt{2}i\omega_{2,+2}\lambda^{\bar{1}}_+
+\sqrt{2}\omega_{2,\bo2}\lambda_-^{1}
+\omega_{2,1\bar{1}}\lambda_-^{\bar{1}}) \nn +
{i\over4}(H_{+-}\lambda^{\bar{1}}_- +2iH_{+\bo}\lambda^{1}_+ +
\sqrt{2}iH_{+2}\lambda^{\bar{1}}_+ +\sqrt{2}H_{\bo2}\lambda^1_- +
H_{1\bar{1}}\lambda^{\bar{1}}_-) -
{\chi\over2}V_IX^I\lambda^{\bar{1}}_- = 0 \ . \nn \end{eqnarray}

\subsection{Constraints on Half-Supersymmetric Solutions}

Substituting the constraints obtained in Section 4, for
quarter-supersymmetric solutions with $\epsilon = \psi^1_+$, back
into the dilatino equations we find

\begin{equation} 2\sqrt{2}\partial_-X^I \lambda_-^1 +4i(F^{I}_{-1} -
X^IH_{-1})\lambda_-^{\bar{1}}-2\sqrt{2}i(F^{I}_{-2} -
X^IH_{-2})\lambda_-^1  = 0 \ , \end{equation}

\bea 8i\chi(X^IV_JX^J - {3\over2}Q^{IJ}V_J)\lambda_+^{\bar{1}} +
2\sqrt{2}\partial_-X^I \lambda_-^{\bar{1}}  \nn + 4i(F^{I}_{-\bo} -
X^IH_{-\bo})\lambda_-^1 + 2\sqrt{2}i(F^{I}_{-2} -
X^IH_{-2})\lambda_-^{\bar{1}} = 0 \ , \eea

\begin{equation} 4\sqrt{2}i\partial_{1}X^I \lambda_-^{\bar{1}} - 4i\partial_2X^I
\lambda_-^1  = 0 \ , \end{equation}

\begin{equation} 4\sqrt{2}i\partial_{\bo}X^I \lambda_-^1
+ 4i\partial_2X^I\lambda_-^{\bar{1}} - 8i\chi(X^IV_JX^J -
{3\over2}Q^{IJ}V_J)\lambda_-^{\bar{1}}= 0 \ . \end{equation}

Substituting the constraints back into the gravitino equations
yields, in the $+$ direction:

\begin{equation} \partial_+\lambda_+^1 - i\omega_{+,-1}\lambda_-^{\bar{1}} +
{\sqrt{2}i\over2}\omega_{+,-2}\lambda_-^1
+{i\over2}\omega_{2,12}\lambda_-^{\bar{1}} +
{\sqrt{2}i\over2}\omega_{-,+2}\lambda_-^1  = 0 \ , \end{equation}

\begin{eqnarray} \partial_{+}\lambda_+^{\bo} +\omega_{+,1\bo}\lambda_+^{\bar{1}}-
i\omega_{+,-\bo}\lambda_-^1
-{\sqrt{2}i\over2}\omega_{+,-2}\lambda_-^{\bo} \nn +
{i\over2}\omega_{2,\bo2}\lambda_-^1 +
{\sqrt{2}i\over6}\omega_{-,+2}\lambda_-^{\bo} -
{\sqrt{2}i\over3}\omega_{1,\bo2}\lambda_-^{\bo} = 0 \ ,
\end{eqnarray}

\begin{equation}  \partial_+\lambda_-^1 = 0 \ , \end{equation}

\begin{equation} (\partial_+ + \omega_{+,1\bo})\lambda_-^{\bo} = 0 \ .\end{equation}

In the $-$ direction:

\begin{equation} \partial_-\lambda_+^1 -i\omega_{-,-1}\lambda_-^{\bo} +
{\sqrt{2}i\over2}\omega_{-,-2}\lambda_-^1 = 0 \ , \end{equation}

\begin{equation} (\partial_- -3i\chi A_-)\lambda_+^{\bar{1}}
 - i\omega_{-,-\bo}\lambda_-^1 -
{\sqrt{2}i\over2}\omega_{-,-2}\lambda_-^{\bo} = 0 \ , \end{equation}

\begin{equation} (\partial_- -2i\chi A_-)\lambda_-^1
-{2\sqrt{2}\over3}\omega_{-,12}\lambda_-^{\bo} -
{2\over3}\omega_{-,1\bo}\lambda_-^1
 = 0 \ , \end{equation}

\begin{equation} (\partial_- - i\chi A_-)\lambda_-^{\bo} +
{2\sqrt{2}\over3}\omega_{-,\bo2}\lambda_-^1 +
{2\over3}\omega_{-,1\bo}\lambda_-^{\bo} +
{\sqrt{2}i\over3}(2\omega_{-,+2} - \omega_{1,\bo2})\lambda_+^{\bo} =
0 \ . \end{equation}

In the $1$ direction:

\begin{eqnarray} \partial_1\lambda_+^1 +
{\sqrt{2}i\over2}\omega_{-,12}\lambda_-^1
+{\sqrt{2}i\over2}\omega_{1,-2}\lambda_-^1 -
i\omega_{1,-1}\lambda_-^{\bo} = 0 \ , \end{eqnarray}

\begin{eqnarray} \partial_1\lambda_+^{\bo}  + \omega_{1,1\bo} \lambda_+^{\bo} -
2\omega_{1,+-}\lambda_+^{\bo} - i\omega_{1,-\bo}\lambda_-^1 +\chi
A_-\lambda_-^1 \nn + {\sqrt{2}i\over6}\omega_{-,12}\lambda_-^{\bo} -
{i\over3}\omega_{-,1\bo}\lambda_-^1 -
{\sqrt{2}i\over2}\omega_{1,-2}\lambda_-^{\bo} = 0 \ , \end{eqnarray}

\begin{equation} (\partial_1 - 2\omega_{1,+-})\lambda_-^1  = 0 \ , \end{equation}

\begin{equation} \partial_1\lambda_-^{\bo} + \omega_{1,1\bo}\lambda_-^{\bo} +
\sqrt{2}\omega_{1,\bo2}\lambda_-^1  = 0 \ . \end{equation}

In the $\bo$ direction:

\begin{eqnarray} \partial_{\bo}\lambda_+^1  + {\sqrt{2}\over3}(2\omega_{-,+2} -
\omega_{1,\bo2})\lambda_+^{\bo}
-{\sqrt{2}i\over6}\omega_{-,\bo2}\lambda_-^1 \nn
+{\sqrt{2}i\over2}\omega_{\bo,-2}\lambda_-^1 -
i\omega_{\bo,-1}\lambda_-^{\bo} +
{i\over3}\omega_{-,1\bo}\lambda_-^{\bo}- \chi A_-\lambda_-^{\bo} = 0
\ , \end{eqnarray}

\begin{equation} \partial_{\bar{1}}\lambda_+^{\bar{1}}
+2\omega_{\bo,+-}\lambda_+^{\bar{1}}  +
\omega_{\bo,1\bar{1}}\lambda_+^{\bar{1}}
-i\omega_{\bo,-\bo}\lambda_-^1 -
{\sqrt{2}i\over2}\omega_{\bo,-2}\lambda_-^{\bar{1}} -
{\sqrt{2}i\over2}\omega_{-,\bo2}\lambda_-^{\bar{1}}= 0
,\end{equation}

\begin{equation} \partial_{\bar{1}}\lambda_-^1  + 2\omega_{\bo,+-}\lambda_-^1
-{2\sqrt{2}\over3}(\omega_{-,+2} +
\omega_{\bo,12})\lambda_-^{\bar{1}} = 0 \ , \end{equation}

\begin{equation} (\partial_{\bar{1}} + \omega_{\bo,1\bo})\lambda_-^{\bar{1}} = 0
\ . \end{equation}

In the $2$ direction:

\begin{equation} \partial_2\lambda_+^1  - \sqrt{2}\lambda_-^1(-\chi A_- +
{i\over3}\omega_{-,1\bo}) - i\omega_{2,-1}\lambda_-^{\bar{1}} +
{\sqrt{2}i\over2}\omega_{2,-2}\lambda_-^1 +
{i\over3}\omega_{-,12}\lambda_-^{\bar{1}} = 0 \ , \end{equation}

\begin{eqnarray} \partial_2\lambda_+^{\bar{1}} + ({2\over3}\omega_{-,+2}
-{1\over3} \omega_{1,\bo2} + \omega_{2,1\bo})\lambda_+^{\bar{1}} -
i\omega_{2,-\bo}\lambda_-^1 + {i\over3}\omega_{-,\bo2}\lambda_-^1
\nn - {\sqrt{2}i\over2}\omega_{2,-2}\lambda_-^{\bar{1}} -
\sqrt{2}(-\chi A_- +{i\over3}\omega_{-,1\bar{1}})\lambda_-^{\bar{1}}
= 0 \ , \end{eqnarray}

\begin{equation} \partial_2\lambda_-^1 - \sqrt{2}\omega_{2,12}\lambda_-^{\bar{1}} =
0 \ , \end{equation}

\begin{equation} \partial_2\lambda_-^{\bar{1}} +
\sqrt{2}\omega_{2,\bo2}\lambda_-^1 -({2\over3}\omega_{-,+2}
-{1\over3} \omega_{1,\bo2} - \omega_{2,1\bo})\lambda_-^{\bar{1}} = 0
\ . \end{equation}

\subsection{Solutions with $\lambda_+^{\alpha} = 0$}

In the case where $\lambda_+^{\alpha} = 0$ we can reduce the
dilatino equations to:

\begin{eqnarray}  \label{d1}  2\sqrt{2}\partial_-X^I \lambda_-^1 +4i(F^{I}_{-1}
- X^IH_{-1})\lambda_-^{\bar{1}}-2\sqrt{2}i(F^{I}_{-2} -
X^IH_{-2})\lambda_-^1  = 0 \ , \end{eqnarray}

\begin{eqnarray} \label{d2} 2\sqrt{2}\partial_-X^I \lambda_-^{\bar{1}}
+4i(F^{I}_{-\bo} - X^IH_{-\bo})\lambda_-^1 + 2\sqrt{2}i(F^{I}_{-2} -
X^IH_{-2})\lambda_-^{\bar{1}} = 0 \ , \end{eqnarray}

\begin{eqnarray} \label{d3} 4\sqrt{2}i\partial_{1}X^I \lambda_-^{\bar{1}} -
4i\partial_2X^I \lambda_-^1  = 0 \ , \end{eqnarray}

\begin{eqnarray} \label{d4} 4\sqrt{2}i\partial_{\bo}X^I \lambda_-^1
+ 4i\partial_2X^I\lambda_-^{\bar{1}} - 8i\chi(X^IV_JX^J -
{3\over2}Q^{IJ}V_J)\lambda_-^{\bar{1}}= 0 \ . \end{eqnarray}

The gravitino equations reduce to, in the $+$ direction:

\begin{eqnarray} \label{+1} i\omega_{+,-1}\lambda_-^{\bar{1}} -
{\sqrt{2}i\over2}\omega_{+,-2}\lambda_-^1
-{i\over2}\omega_{2,12}\lambda_-^{\bar{1}} -
{\sqrt{2}i\over2}\omega_{-,+2}\lambda_-^1 = 0 \ , \end{eqnarray}

\begin{eqnarray} \label{+2} i\omega_{+,-\bo}\lambda_-^1
+{\sqrt{2}i\over2}\omega_{+,-2}\lambda_-^{\bar{1}} -
{i\over2}\omega_{2,\bo2}\lambda_-^1 -
{\sqrt{2}i\over6}\omega_{-,+2}\lambda_-^{\bar{1}} +
{\sqrt{2}i\over3}\omega_{1,\bo2}\lambda_-^{\bar{1}} = 0 \ ,
\end{eqnarray}

\begin{eqnarray}  \partial_+\lambda_-^1 = 0 \ , \end{eqnarray}

\begin{eqnarray} (\partial_+ + \omega_{+,1\bar{1}})\lambda_-^{\bar{1}} = 0 \ .
\end{eqnarray}

In the $-$ direction:

\begin{eqnarray} \label{-1} i\omega_{-,-1}\lambda_-^{\bar{1}} -
{\sqrt{2}i\over2}\omega_{-,-2}\lambda_-^1 = 0 , \end{eqnarray}

\begin{eqnarray} \label{-2} i\omega_{-,-\bo}\lambda_-^1 +
{\sqrt{2}i\over2}\omega_{-,-2}\lambda_-^{\bar{1}} = 0 \ ,
\end{eqnarray}

\begin{eqnarray} (\partial_- -2i\chi A_-)\lambda_-^1
-{2\sqrt{2}\over3}\omega_{-,12}\lambda_-^{\bar{1}} -
{2\over3}\omega_{-,1\bar{1}}\lambda_-^1
 = 0 \ , \end{eqnarray}

\begin{eqnarray} (\partial_- - i\chi A_-)\lambda_-^{\bar{1}} +
{2\sqrt{2}\over3}\omega_{-,\bo2}\lambda_-^1 +
{2\over3}\omega_{-,1\bar{1}}\lambda_-^{\bar{1}}  = 0 \ .
\end{eqnarray}

In the $1$ direction:

\begin{eqnarray} \label{11} {\sqrt{2}i\over2}\omega_{-,12}\lambda_-^1
+{\sqrt{2}i\over2}\omega_{1,-2}\lambda_-^1 -
i\omega_{1,-1}\lambda_-^{\bar{1}} = 0 \ , \end{eqnarray}

\begin{eqnarray} \label{12} i\omega_{1,-\bo}\lambda_-^1 - \chi A_-\lambda_-^1
-{\sqrt{2}i\over6}\omega_{-,12}\lambda_-^{\bar{1}} +
{i\over3}\omega_{-,1\bar{1}}\lambda_-^1 +
{\sqrt{2}i\over2}\omega_{1,-2}\lambda_-^{\bar{1}} = 0 \ ,
\end{eqnarray}

\begin{eqnarray} (\partial_1 - 2\omega_{1,+-})\lambda_-^1  = 0 \ , \end{eqnarray}

\begin{eqnarray} \partial_1\lambda_-^{\bar{1}} +
\omega_{1,1\bar{1}}\lambda_-^{\bar{1}} +
\sqrt{2}\omega_{1,\bo2}\lambda_-^1  = 0 \ . \end{eqnarray}

In the $\bar{1}$ direction:

\begin{eqnarray} \label{1bar1} {\sqrt{2}i\over6}\omega_{-,\bo2}\lambda_-^1
-{\sqrt{2}i\over2}\omega_{\bo,-2}\lambda_-^1 +
i\omega_{\bo,-1}\lambda_-^{\bar{1}} -
{i\over3}\omega_{-,1\bo}\lambda_-^{\bar{1}}+ \chi
A_-\lambda_-^{\bar{1}} = 0 \ , \end{eqnarray}

\begin{eqnarray} \label{1bar2} i\omega_{\bo,-\bo}\lambda_-^1 +
{\sqrt{2}i\over2}\omega_{\bo,-2}\lambda_-^{\bar{1}} +
{\sqrt{2}i\over2}\omega_{-,\bo2}\lambda_-^{\bar{1}}= 0 \ ,
\end{eqnarray}

\begin{eqnarray} \partial_{\bar{1}}\lambda_-^1  + 2\omega_{\bo,+-}\lambda_-^1 -
{2\sqrt{2}\over3}(\omega_{-,+2} +
\omega_{\bo,12})\lambda_-^{\bar{1}} = 0 \ , \end{eqnarray}

\begin{eqnarray} (\partial_{\bar{1}} + \omega_{\bo,1\bo})\lambda_-^{\bar{1}} =
0 \ .
\end{eqnarray}

In the $2$ direction:

\begin{eqnarray} \label{21} \sqrt{2}\lambda_-^1(-\chi A_- -
{i\over3}\omega_{-,1\bo}) + i\omega_{2,-1}\lambda_-^{\bar{1}} -
{\sqrt{2}i\over2}\omega_{2,-2}\lambda_-^1 -
{i\over3}\omega_{-,12}\lambda_-^{\bar{1}} = 0 \ , \end{eqnarray}

\begin{eqnarray} \label{22} i\omega_{2,-\bo}\lambda_-^1 +
{i\over3}\omega_{-,\bo2}\lambda_-^1 -
{\sqrt{2}i\over2}\omega_{2,-2}\lambda_-^{\bar{1}} - \sqrt{2}(-\chi
A_- + {i\over3}\omega_{-,1\bo})\lambda_-^{\bar{1}} = 0 \ ,
\end{eqnarray}

\begin{eqnarray} \partial_2\lambda_-^1 - \sqrt{2}\omega_{2,12}\lambda_-^{\bar{1}}
= 0 \ , \end{eqnarray}

\begin{eqnarray} \partial_2\lambda_-^{\bar{1}} +
\sqrt{2}\omega_{2,\bo2}\lambda_-^1 -({2\over3}\omega_{-,+2}
-{1\over3} \omega_{1,\bo2} - \omega_{2,1\bo})\lambda_-^{\bar{1}} = 0
\ .
\end{eqnarray}

\medskip

\acknowledgments

Jai Grover thanks the Cambridge Commonwealth Trusts for support.
Jan Gutowski thanks Hari Kunduri for useful discussions.
The work of W. Sabra was supported in part by the National Science
Foundation under grant number PHY-0703017.


\end{document}